\documentclass[twocolumn,twocolappendix]{aastex701} %,linenumbers

\usepackage{amsmath}
\usepackage{booktabs, makecell, longtable}
\usepackage{multirow}

\newcommand{\aco}{$\alpha_\mathrm{CO}$}
\newcommand{\Ngal}{62~}
\defcitealias{co-to-h2}{B13}

\shorttitle{SFE and Galaxy Quenching with GBT-EDGE}
\shortauthors{Teng et al.}
\begin{document}

\title{The EDGE-CALIFA Survey: Star Formation Efficiency and Galaxy Quenching across \Ngal Main Sequence, Green Valley, and Red Galaxies}

\newcommand{\Alberta}{\affiliation{Department of Physics, University of Alberta, Edmonton, AB T6G 2E1, Canada}}

\newcommand{\ASIAA}{\affiliation{Institute of Astronomy and Astrophysics, Academia Sinica, No. 1, Sec. 4, Roosevelt Road, Taipei 10617, Taiwan}}

\newcommand{\Bonn}{\affiliation{Argelander-Institut f\"ur Astronomie, University of Bonn, Auf dem H\"ugel 71, 53121 Bonn, Germany}}

\newcommand{\Heidelberg}{\affiliation{Astronomisches Rechen-Institut, Zentrum f\"{u}r Astronomie der Universit\"{a}t Heidelberg, M\"{o}nchhofstra\ss e 12-14, D-69120 Heidelberg, Germany}}

\newcommand{\IAC}{\affiliation{Instituto de Astrof\'\i sica de Canarias, La Laguna, Tenerife, E-38200, Spain}}

\newcommand{\ITA}{\affiliation{Universit\"{a}t Heidelberg, Zentrum f\"{u}r Astronomie, Institut f\"{u}r Theoretische Astrophysik, \\ Albert-Ueberle-Str 2, D-69120 Heidelberg, Germany}}

\newcommand{\Leiden}{\affiliation{Leiden Observatory, Leiden University, P.O. Box 9513, 2300 RA Leiden, The Netherlands}}

\newcommand{\Maryland}{\affiliation{Department of Astronomy, University of Maryland, 4296 Stadium Drive, College Park, MD 20742, USA}}

\newcommand{\MPE}{\affiliation{Max-Planck-Institut f\"{u}r extraterrestrische Physik, Giessenbachstra{\ss}e 1, D-85748 Garching, Germany}}

\newcommand{\MPIA}{\affiliation{Max-Planck-Institut f\"{u}r Astronomie, K\"{o}nigstuhl 17, D-69117, Heidelberg, Germany}}

\newcommand{\NRAO}{\affiliation{National Radio Astronomy Observatory, 520 Edgemont Road, Charlottesville, VA 22903-2475, USA}}

\newcommand{\OSU}{\affiliation{Department of Astronomy, The Ohio State University, 140 West 18th Avenue, Columbus, OH 43210, USA}}

\newcommand{\STScI}{\affiliation{Space Telescope Science Institute, 3700 San Martin Drive, Baltimore, MD 21218, USA}}

\newcommand{\Toledo}{\affiliation{Department of Physics and Astronomy, University of Toledo, Ritter Obs., MS \#113, Toledo, OH 43606, USA}}

\newcommand{\UCSD}{\affiliation{Department of Astronomy \& Astrophysics, University of California San Diego, 9500 Gilman Drive, La Jolla, CA 92093, USA}}

\newcommand{\UIUC}{\affiliation{Department of Astronomy, University of Illinois at Urbana-Champaign, 1002 W. Green Street, Urbana, IL 61801, USA}}

\newcommand{\UNAMe}{\affiliation{Universidad Nacional Aut\'onoma de M\'exico, Instituto de Astronom\'ia, AP 106, Ensenada 22800, BC, M\'exico}}

% please add you author info below.

\correspondingauthor{Yu-Hsuan Teng}

\author[0000-0003-4209-1599]{Yu-Hsuan Teng}
\Maryland
\email[show]{yhteng@umd.edu}

\author[0000-0002-5480-5686]{Alberto D. Bolatto}
\Maryland
\affiliation{Joint Space-Science Institute, University of Maryland, College Park, MD 20742, USA}
\email{bolatto@umd.edu}

\author[0000-0003-1774-3436]{Peter J. Teuben}
\Maryland
\email{teuben@umd.edu}

\author[0000-0002-5204-2259]{Erik Rosolowsky}
\Alberta
\email{rosolowsky@ualberta.ca}

\author[0000-0003-1924-1122]{David T. Frayer}
\affiliation{Green Bank Observatory, 155 Observatory Road, Box 2, Green Bank, WV 24944, USA}
\email{dfrayer@nrao.edu}

\author[0000-0002-3227-4917]{Amanda A. Kepley}
\NRAO
\email{akepley@nrao.edu}

\author[0000-0001-6444-9307]{Sebastian F. Sanchez}
\UNAMe
\IAC
\email{sfsanchez@astro.unam.mx}

\author[0000-0002-7759-0585]{Tony Wong}
\UIUC
\email{wongt@illinois.edu}

\author[0000-0002-2545-1700]{Adam K. Leroy}
\OSU
\email{leroy.42@osu.edu}

\author[0000-0001-6498-2945]{Dario Colombo}
\Bonn
\email{dcolombo@uni-bonn.de}

\author[0000-0002-9511-1330]{Serena A. Cronin}
\Maryland
\email{cronin@umd.edu}

\author[0000-0002-4235-7337]{K. Decker French}
\UIUC
\email{deckerkf@illinois.edu}

\author[0000-0002-2262-5875]{Veselina Kalinova}
\affiliation{Max Planck Institute for Radio Astronomy, Auf dem Hügel 69, 53121 Bonn,
Germany}
\affiliation{Institute of Astronomy and National Astronomical Observatory,
Bulgarian Academy of Sciences, 72 Tsarigradsko Chaussee Blvd., 1784 Sofia,
Bulgaria}
\email{kalinova@mpifr-bonn.mpg.de}

\author[0000-0003-2508-2586]{Rebecca C. Levy}
\STScI
\email{rlevy.astro@gmail.com}

\author[0000-0002-4378-8534]{Karin M. Sandstrom}
\UCSD
\email{kmsandstrom@ucsd.edu}

\author[0000-0002-5877-379X]{Vicente Villanueva}
\affiliation{Instituto de Estudios Astrofísicos, Facultad de Ingeniería y Ciencias, Universidad Diego Portales, Av. Ejército Libertador 441, 8370191 Santiago, Chile}
\affiliation{Millennium Nucleus for Galaxies, MINGAL}
\email{vicente.avl365@gmail.com}

\author[0000-0003-2405-7258]{Jorge K. Barrera-Ballesteros}
\UNAMe
\email{jkbarrerab@astro.unam.mx}

\author[0009-0001-1221-0975]{Zein Bazzi}
\Bonn
\email{zbazzi@uni-bonn.de}

\author[0000-0001-5301-1326]{Yixian Cao}
\MPE
\email{ycao@mpe.mpg.de}

\author[0000-0002-8432-3362]{Alex Green}
\UIUC
\email{alexg4@illinois.edu}

\author[0000-0002-2775-0595]{Rodrigo Herrera-Camus}
\affiliation{Departamento de Astronomía, Universidad de Concepción, Concepción, Chile}
\affiliation{Millennium Nucleus for Galaxies, MINGAL}
\email{rhc@udec.cl} 

\author[0000-0001-7231-7953]{Eduardo A. D. Lacerda}
\address{Instituto de Astronom\'ia, Universidad Nacional Aut\'onoma de M\'exico, AP 70-264, CDMX 04510, M\'exico}
\email{dhubax@gmail.com} 

\author[0000-0003-0665-6505]{Jialu Li}
\Maryland
\email{jialu@umd.edu}

\author[0000-0001-9226-9178]{Alejandra Z. Lugo-Aranda}
\UNAMe
\email{alugo@astro.unam.mx} 

\author[0009-0008-1239-9959]{Jocabed Martinez-Lopez}
\affiliation{Departamento de Astronomía, Universidad de Concepción, Concepción, Chile}
\affiliation{Millennium Nucleus for Galaxies, MINGAL}
\email{jocmartinez2023@udec.cl}

\author[0000-0003-1356-1096]{Elizabeth Tarantino}
\STScI
\email{etarantino@stsci.edu} 

\author[0000-0002-6582-4946]{Akshat Tripathi}
\UIUC
\email{akshatt3@illinois.edu}

\author[0009-0007-8404-6282]{Carolyn G. Volpert}
\Maryland
\email{cvolpert@astro.umd.edu}

\author[0000-0003-2812-8607]{Di Wen}
\affiliation{Kapteyn Astronomical Institute, University of Groningen, P.O. Box 800, 9700AV Groningen, The Netherlands}
\email{wen@astro.rug.nl}

\begin{abstract}

We present GBT-EDGE, a new CO\,(1--0) survey using the Green Bank Telescope to map \Ngal nearby (10--140~Mpc) galaxies spanning the star-forming main sequence (SFMS), green valley, and red sequence. The galaxy sample is selected from the CALIFA survey with integral field spectroscopy (IFS), which provides a representative census of local galactic environments. Combining the CO dataset with CALIFA's optical IFS measurements, we derive molecular gas masses, star formation rates (SFR), metallicities, and stellar mass densities to measure star formation efficiency (SFE) and investigate the physical drivers of galaxy quenching.
We obtain a median molecular gas depletion time of $2.10^{+2.35}_{-1.31}$, $6.90^{+17.00}_{-3.67}$, and $127.7^{+201.6}_{-113.4}$ Gyr for our sample of main sequence, green valley, and red galaxies, respectively, assuming a Galactic CO-to-H$_2$ conversion factor.
By applying various conversion factor prescriptions, we also confirm a systematic decrease of SFE with galaxy's offset below the SFMS, regardless of the adopted prescription. 
This suggests that the low SFR in some quenched galaxies is primarily driven by suppressed SFE rather than an absence of molecular gas. 
Our results provide evidence that galaxies below the main sequence can retain substantial molecular gas reservoirs comparable to star-forming galaxies, but they exhibit longer depletion times and form stars inefficiently, possibly due to the combined effects of low gas density and morphological quenching mechanisms. 

\end{abstract}


\section{Introduction \label{sec:intro}} 

Galaxy evolution is driven by variations in the level of star formation activities within galaxies. 
Observational studies on nearby galaxies over the past two decades have confirmed a strong correlation between a galaxy's star formation rate (SFR) and its stellar mass ($M_\mathrm{star}$), namely, the star formation main sequence \citep[SFMS;][]{2004MNRAS.351.1151B,2007ApJS..173..267S,2016ApJ...821L..26C,2016MNRAS.462.1749S}. This relation has also been found to hold for spatially resolved observations down to kpc scales, and thus the relation is sometimes expressed in terms of surface densities \citep[$\Sigma_\mathrm{SFR}$ and $\Sigma_\mathrm{star}$; e.g.,][]{2019ApJ...884L..33L,2021MNRAS.503.1615S}. 
The established SFMS relation has further led to a common galaxy classification based on the offset from the SFMS relation: main sequence (MS) galaxies are those aligned with the SFMS, red galaxies (RGs) are those falling on the ``red cloud sequence'' which is substantially below the SFMS (e.g., \citealt{2007ApJS..173..293W}), and green valley (GV) galaxies are those in between, which implies a transitioning stage where star formation begins to shut down \citep{2007ApJS..173..267S}. 
As a result, the offset from the SFMS (or many studies also use ``specific SFR'', which is defined as sSFR $\equiv$ SFR/$M_\mathrm{star}$) has been widely adopted to indicate a galaxy's evolutionary stage in terms of its star formation level \citep[e.g.,][]{2017ApJS..233...22S,2020A&A...644A..97C,2023ApJ...958..183S}.

To probe the reasons for the rise and fall of star formation in galaxies, it is critical to study the cold molecular interstellar medium, which is the birthplace of stars and thus determines the capacity for star formation in a galaxy. 
Specifically, star formation is governed by the amount of molecular gas as well as how efficiently molecular gas is converted into stars \citep{2012ARA&A..50..531K,2022ARA&A..60..319S}.
Previous studies have revealed a roughly constant star formation efficiency (SFE $\equiv$ SFR/$M_\mathrm{mol}$, where $M_\mathrm{mol}$ represents the molecular gas mass) in nearby MS galaxy disks, where the molecular gas depletion time ($t_\mathrm{dep} \equiv \mathrm{SFE}^{-1}$) is around 2~Gyr, despite variations found between galaxy centers \citep[e.g.,][]{2008AJ....136.2846B,2011MNRAS.415...61S,2013AJ....146...19L,2017ApJ...849...26U,2019PASJ...71S..15M,2023ApJ...945L..19S,2024ApJ...961...42T}.
The SFE in star-forming galaxy centers is sometimes found to be higher than the value in their disks, which would enhance star formation and support central starbursts \citep[e.g.,][]{2017ApJ...849...26U,2020MNRAS.492.6027E}. 
SFE in centers may also be lower than in the disks, which can indicate galaxy quenching due to feedback from active galactic nuclei (AGN; \citealt{2012Natur.485..213P}) or morphological quenching driven by bars or bulges \citep{2009ApJ...707..250M,2012ApJ...758...73S,2018MNRAS.475.1791C,2023ApJ...943....7M}.   
While these SFE estimations are known to be sensitive to the choices of SFR indicator and CO-to-H$_2$ conversion factor (\aco; \citealt{co-to-h2}), measurements and/or prescriptions for SFR and \aco~are relatively well-developed for MS galaxies \citep[e.g.,][]{2023ApJ...950..119T,2024ApJ...961...42T,2024ApJ...964...18C,2024ARA&A..62..369S,2025ApJ...994..263S}, and thus recent SFE studies are able to account for such systematic variations \citep[e.g.,][]{2023ApJ...945L..19S,2024A&A...687A.293Q,2024ApJ...961...42T,2025A&A...699A.367C}.   

On the other hand, SFE across quenched galaxies, including GV galaxies, and especially RGs, remains underexplored. This is because obtaining high-quality molecular gas observations (usually via CO rotational line emission) in such environments requires significant integration time, given the sensitivity of current millimeter-wave telescopes. 
Therefore, current molecular gas studies are limited to a small sample of GV galaxies and tend to focus only on the transition from MS to GV rather than extending to a comparable sample of RGs. 
The sensitivity limitation also makes measuring accurate gas mass and \aco\ even more challenging in the GV and RGs, and thus most studies on these systems still rely on assuming a constant Galactic-like \aco\ value to determine molecular gas-related properties \citep[e.g.,][]{2013MNRAS.432.1796A,2014MNRAS.444.3427D,2022ApJ...926..175L,2024ApJ...962...88V}.
Furthermore, while both molecular gas fraction and SFE are known to play a role in galaxy quenching, there is not yet a consensus about whether one of these factors dominates, and how they might change across different galaxy evolutionary stages \citep{2017ApJS..233...22S,2020A&A...644A..97C,2020MNRAS.498L..66B,2020MNRAS.493L..39E,2021MNRAS.500.2289K,2022ApJ...926..175L,2024ApJ...964..120P,2024ApJ...962...88V}. 

The physical cause of a reduced molecular gas fraction or SFE in quenched galaxies is also under debate.
Previous studies have found that galaxies below the MS tend to be associated with early-type morphologies (e.g., bigger bulges; \citealt{2019MNRAS.488.3929C}) and they are more likely to host AGN \citep[e.g.,][]{2019MNRAS.484.4360A,2020MNRAS.492.3073L}. Such structural differences can affect the local star formation processes but also complicate explanations of different quenching mechanisms. For instance, the existence of AGN could either indicate increased gas availability due to black hole accretion and its co-evolution with the galaxy \citep{2014ARA&A..52..589H,2022MNRAS.514.2936W}, or it may lead to a strong depletion of gas via AGN feedback processes \citep{2017ApJS..233...22S,2021MNRAS.505L..46E}. Similarly, bar-driven inflows are known to bring in more gas and boost star formation \citep{1999ApJ...525..691S,2005ApJ...632..217S,2007PASJ...59..117K,2019MNRAS.484.5192C,2020MNRAS.499.4455T,2022A&A...666A.175Y}, whereas the gas stabilization from bulge potential or bar-induced shear can support the presence of substantial quiescent gas reservoirs \citep{2009ApJ...707..250M,2014MNRAS.444.3427D,2020MNRAS.495..199G,2023ApJ...943....7M}. 

In this paper, we present new CO $J$=1--0 observations with the Robert C. Byrd Green Bank Telescope (GBT) toward \Ngal nearby galaxies with $M_\mathrm{star}$ between $3\times10^8$ to $3\times10^{11}$~M$_\odot$. Our targets include GV and RGs in a similar sample size ($\sim$10 galaxies each), along with a sizable sample of MS galaxies to be compared with. With optical integral field unit (IFU) data available from the Calar Alto Legacy Integral Field Area (CALIFA) survey \citep{2012A&A...538A...8S,2016A&A...594A..36S}, this sample allows a comprehensive study of SFE variations across galaxy populations with varying locations in the SFMS relation. We estimate $M_\mathrm{mol}$ and SFE with multiple \aco\ treatments and investigate their roles in the quenching of star formation throughout different galaxy evolutionary stages.

The paper is structured as follows. Section~\ref{sec:data} describes our sample, observations, data processing, and ancillary data. Section~\ref{sec:result} presents our analyses and results, including the derivation of data products and various galaxy-integrated properties. In Section~\ref{sec:discussion}, we discuss implications from our results on galaxy quenching mechanisms and compare them to those from the literature. The conclusions are summarized in Section~\ref{sec:conclusion}.
In this work, we assume a flat $\Lambda$CDM cosmology with $H_0 = 70$ km~s$^{-1}$~Mpc$^{-1}$, $\Omega_\mathrm{m}$ = 0.27, and $\Omega_\mathrm{\Lambda}$ = 0.73. 

\section{Observations and Data  \label{sec:data}}

\begin{figure}
\centering
%\vspace{3ex}
\includegraphics[width=\linewidth]{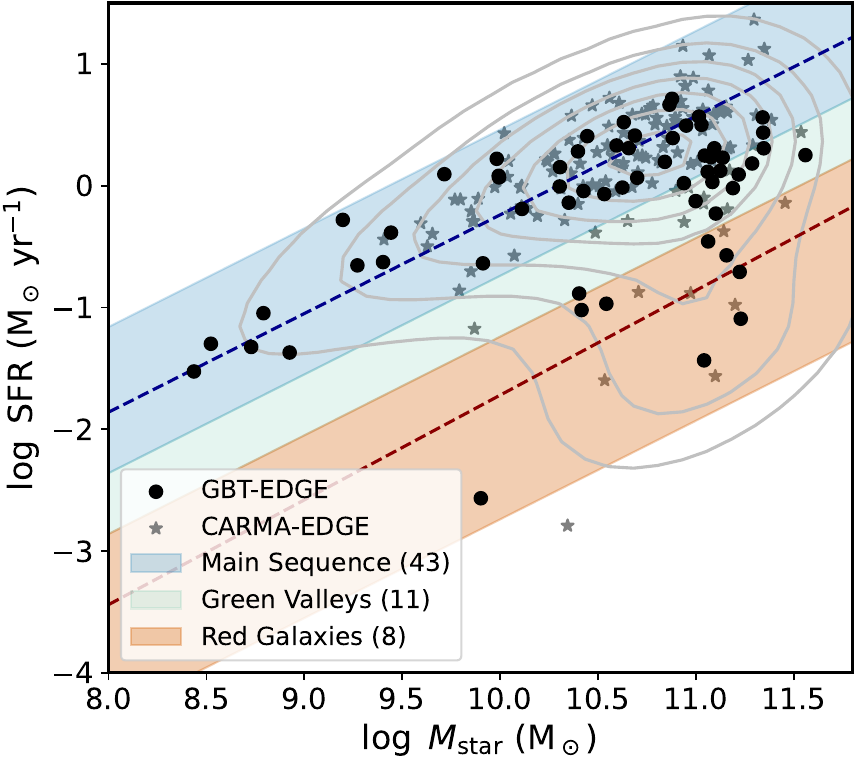}\\
\caption{The relation between global star formation rate (SFR) and stellar mass ($M_\mathrm{star}$) for our GBT sample (black points; \Ngal galaxies), the CARMA sample \citep[gray stars;][]{2017ApJ...846..159B}, and the full CALIFA sample \citep[contours;][]{2016A&A...594A..36S}. The blue (upper) and red (lower) dashed lines indicate best linear fits from \citet{2016ApJ...821L..26C} for the main sequence and red galaxies, respectively. The GBT-EDGE sample is representative of the $z=0$ galaxy population with $M_\mathrm{star} = 10^{8.5-11.5}$ M$_\odot$.}
\label{fig:sfms}
\end{figure}

\begin{figure*}
\centering
\includegraphics[width=\linewidth]{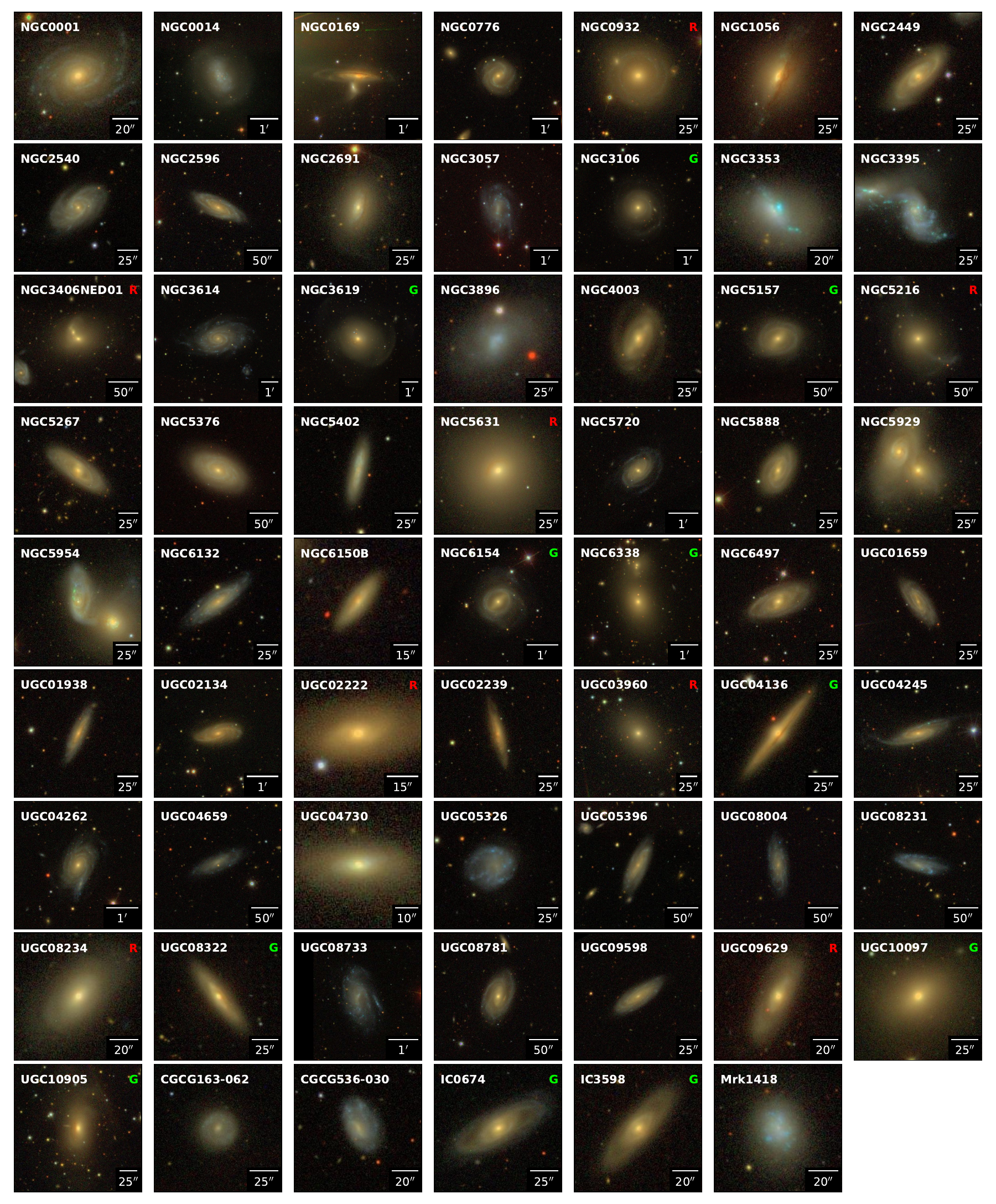}\\
\caption{SDSS \textit{g} (blue channel), \textit{r} (green channel), and \textit{i} (red channel) composite images for all \Ngal galaxies in our GBT-EDGE sample. Green valley (GV) and red galaxies (RGs) are labeled by a letter `G' and `R', respectively, in the top-right corners of their panels. These galaxies cover a variety of morphologies and galaxy environments, which constitute a representative sample of the local Universe (Figure~\ref{fig:sfms}) and enable comprehensive studies on the galaxy quenching process.}
\label{fig:sdss}
\end{figure*}

\subsection{The GBT-EDGE Sample  \label{subsec:gbt_sample}} 

We use the GBT to observe CO\,(1--0) across \Ngal galaxies selected from the CALIFA survey \citep{2012A&A...538A...8S}, aiming to complement previous Extragalactic Database for Galaxy Evolution (EDGE) surveys \citep{2017ApJ...846..159B,2020A&A...644A..97C,2024ApJ...962...88V,2024ApJS..271...35W}.
The selection criteria of our parent sample, the CALIFA galaxies, include redshift ($0.03 > z > 0.005$), Galactic latitude ($|b| > 20^\circ$), declination ($> 7^\circ$), and cuts in the angular isophotal diameter ($45''-79.2''$) and flux ($< 20$ in Petrosian magnitudes) in the SDSS $r$-band images \citep{2014A&A...569A...1W,2016A&A...594A..36S}. 
We note that the cut on low redshift effectively excluded objects with distance $\lesssim 20$~Mpc, which avoided the sample to be overwhelmed by dwarf galaxies and/or compact objects in the local Universe. In addition, the imposed flux limit excluded many low surface brightness objects. These selection criteria also indicate that the largest and most massive galaxies in the sample are more distant than the smallest and least massive galaxies.
To select our GBT-EDGE sample from CALIFA, we further impose a declination cut of $\delta > 10^\circ$ and avoid any overlap with the published CARMA-EDGE sample \citep{2017ApJ...846..159B}. This results in our proposed sample of 150 galaxies for GBT observations. However, 79 of the galaxies were not observed and 9 of the observed galaxies were further removed due to severe artifacts, leaving us with the final sample of \Ngal galaxies (see Section~\ref{subsec:observation} for more details).

Figure~\ref{fig:sfms} shows the SFR--$M_\mathrm{star}$ (or SFMS) relation of the \Ngal observed galaxies, with comparison to the full CALIFA sample and a previously published sub-sample using the Combined Array for Research in Millimeter-wave Astronomy (CARMA; \citealt{2017ApJ...846..159B}). 
Our sample of galaxies is representative of the galaxy population with $M_\mathrm{star} = 10^{8.5-11.5}$ M$_\odot$ in the local Universe, spanning distances from 10--140~Mpc (see Table~\ref{tab:sample}). 
Figure~\ref{fig:sdss} presents the optical composite images of these galaxies from the Sloan Digital Sky Survey (SDSS), which show diverse morphological and environmental conditions. The basic properties of the entire galaxy sample are listed in Table~\ref{tab:sample}.

Our GBT sample selection does not overlap with that of the CARMA observations and extends to lower $M_\mathrm{star}$, covering 43 galaxies on the main sequence, 11 galaxies in the green valley, and 8 galaxies on the red cloud. In this work, we define main sequence (MS) galaxies as those with SFRs between -0.5 and 0.7~dex from the best-fit relation in \citet{2016ApJ...821L..26C}:
\begin{equation}
\log(\mathrm{SFR})_\mathrm{MS} = 0.81 \log(M_\mathrm{star}) - 8.34~,
\label{eqn_sfms}
\end{equation}
where $M_\mathrm{star}$ and SFR are in units of M$_\odot$ and M$_\odot$~yr$^{-1}$, respectively. We also define GV as galaxies with SFRs between -0.5 and -1~dex from Equation~\ref{eqn_sfms}. Similarly, RGs are defined as those having SFR $<-1$~dex from Equation~\ref{eqn_sfms}. These boundary definitions are consistent with previous studies showing similar ranges for these galaxy populations \citep{2016ApJ...821L..26C,2018RMxAA..54..217S,2020A&A...644A..97C,2020ApJ...903..145L,2024ApJ...962...88V}.

\begin{longtable*}{lcccccccccc}
\caption{Basic Properties of the GBT-EDGE Sample \label{tab:sample}} \\
\hline \hline
 Galaxy        & R.A.        & Decl.       & Dist.    & log($d_{25}$) & P.A.   & Incl. & vmaxg & log(SFR)       & log($M_\mathrm{star}$)      & Type \\
 &  [deg] & [deg] & [Mpc] & [0.1~arcmin] & [deg] & [deg] & [km~s$^{-1}$] & [M$_\odot$~yr$^{-1}$] & [M$_\odot$] & \\
   (1) & (2) & (3) & (4) & (5) & (6) & (7) & (8) & (9) & (10) & (11)  \\ 
\hline
\endfirsthead

\toprule
 Galaxy        & R.A.        & Decl.       & Dist.    & log($d_{25}$) & P.A.   & Incl. & vmaxg & log(SFR)       & log($M_\mathrm{star}$)      & Type \\
 &  [deg] & [deg] & [Mpc] & [0.1~arcmin] & [deg] & [deg] & [km~s$^{-1}$] & [M$_\odot$~yr$^{-1}$] & [M$_\odot$] & \\
   (1) & (2) & (3) & (4) & (5) & (6) & (7) & (8) & (9) & (10) & (11)  \\ 
\midrule
\endhead
\midrule
\multicolumn{11}{r}{{Continued on next page}} \\
\endfoot

\bottomrule \\
\multicolumn{11}{l}{\parbox{\dimexpr\textwidth-2\tabcolsep}{%
     \textbf{Note.} (1) Galaxy name; (2, 3) central position of the galaxy in J2000 coordinates; (4) luminosity distance \citep{2016RMxAA..52..171S}; (5--8) log of apparent diameter in 0.1~arcmin, position angle, inclination, and apparent maximum rotation velocity of gas, all from HyperLeda\textsuperscript{\ref{hyperleda}} \citep{2014A&A...570A..13M}; (9, 10) log of star formation rate and stellar mass \citep{2016A&A...594A..36S,2024ApJS..271...35W}; (11) galaxy classification as main sequence (MS), green valley (GV), or red galaxy (RG), as defined in Figure~\ref{fig:sfms} via Equation~\ref{eqn_sfms}.}} \\
\endlastfoot

NGC0001      & 1.816000   & 27.708250 & 64.1  & 1.2    & 110.5 & 48.8  & 146   & 0.71   & 10.88  & MS   \\
NGC0014      & 2.192375   & 15.815750 & 12.0  & 1.18   & 23.1  & 55.7  & 35.2  & -1.32  & 8.73   & MS   \\
NGC0169      & 9.215500   & 23.990361 & 65.3  & 1.18   & 92.5  & 69.8  & 278   & 0.56   & 11.34  & MS   \\
NGC0776      & 29.977083  & 23.644389 & 69.4  & 1.13   & 50.0  & 18.3  & 60    & 0.57   & 11.02  & MS   \\
NGC0932      & 36.977833  & 20.332500 & 57.7  & 1.3    & 40.0  & 24.4  & 90    & -0.46  & 11.06  & RG   \\
NGC1056      & 40.701625  & 28.574111 & 21.9  & 1.27   & 162.5 & 46.7  & 124.2 & 0.15   & 10.31  & MS   \\
NGC2449      & 116.834542 & 26.930306 & 69.6  & 1.16   & 135.5 & 73.7  & 101   & 0.02   & 10.94  & MS   \\
NGC2540      & 123.193500 & 26.361833 & 89.4  & 1.094  & 123.3 & 60.4  & 153   & 0.52   & 10.63  & MS   \\
NGC2596      & 126.860250 & 17.283972 & 84.5  & 1.14   & 63.9  & 74.2  & 203.2 & 0.67   & 10.87  & MS   \\
NGC2691      & 133.693083 & 39.538778 & 56.1  & 1.1    & 166.0 & 47.9  & 182   & 0.33   & 10.59  & MS   \\
NGC3057      & 151.416583 & 80.284972 & 21.2  & 1.16   & 8.0   & 58.3  & 100   & -0.65  & 9.27   & MS   \\
NGC3106      & 151.021833 & 31.185472 & 88.8  & 0.95   & 150.0 & 26.0  & 102   & 0.09   & 11.22  & GV   \\
NGC3353      & 161.343042 & 55.960306 & 13.6  & 1.13   & 75.7  & 45.5  & 41.4  & -0.28  & 9.20   & MS   \\
NGC3395      & 162.458792 & 32.982889 & 22.3  & 1.2    & 40.5  & 57.8  & 86.1  & 0.09   & 9.72   & MS   \\
NGC3406NED01 & 162.932542 & 51.023083 & 107.0 & 1.16   & 84.4  & 62.3  & 150   & -0.71  & 11.22  & RG   \\
NGC3614      & 169.588833 & 45.747972 & 32.7  & 1.4    & 87.7  & 47.0  & 131.5 & -0.19  & 10.11  & MS   \\
NGC3619      & 169.839792 & 57.758111 & 21.8  & 1.59   & 25.0  & 42.4  & 165.1 & -0.89  & 10.40  & GV   \\
NGC3896      & 177.234792 & 48.674639 & 12.7  & 1.15   & 128.4 & 55.4  & 110.6 & -1.53  & 8.44   & MS   \\
NGC4003      & 179.495958 & 23.124917 & 93.5  & 1.02   & 159.7 & 65.2  & 150   & 0.26   & 11.08  & MS   \\
NGC5157      & 201.820208 & 32.030778 & 105.0 & 1.12   & 115.5 & 35.4  & 241   & -0.23  & 11.10  & GV   \\
NGC5216      & 203.028625 & 62.700694 & 41.3  & 1.23   & 54.0  & 83.9  & 150.1 & -1.02  & 10.42  & RG   \\
NGC5267      & 205.166417 & 38.794194 & 84.8  & 1.19   & 53.3  & 75.1  & 69.8  & 0.23   & 11.08  & MS   \\
NGC5376      & 208.816875 & 59.506611 & 29.5  & 1.17   & 66.3  & 54.8  & 135   & -0.14  & 10.35  & MS   \\
NGC5402      & 209.569000 & 59.815000 & 42.8  & 1.05   & 166.0 & 81.2  & 122.2 & 0.07   & 9.99   & MS   \\
NGC5631      & 216.638542 & 56.582528 & 27.4  & 1.29   & 150.0 & 0.0   & 165.4 & -0.97  & 10.54  & RG   \\
NGC5720      & 219.638625 & 50.815222 & 111.0 & 1.28   & 132.0 & 52.0  & 194.8 & 0.31   & 11.09  & MS   \\
NGC5888      & 228.280500 & 41.264694 & 125.0 & 1.12   & 156.7 & 54.6  & 150   & 0.44   & 11.34  & MS   \\
NGC5929      & 231.526625 & 41.671667 & 35.3  & 0.96   & 38.2  & 24.4  & 111.8 & 0.06   & 10.70  & MS   \\
NGC5954      & 233.645500 & 15.200167 & 27.0  & 1.01   & 19.2  & 63.6  & 111.4 & 0.22   & 9.98   & MS   \\
NGC6132      & 245.911708 & 11.787194 & 71.0  & 1.11   & 126.5 & 79.9  & 154.6 & 0.28   & 10.40  & MS   \\
NGC6150B     & 246.435167 & 40.475667 & 137.0 & 0.96   & 142.9 & 78.8  & 150   & 0.50   & 11.03  & MS   \\
NGC6154      & 246.377000 & 49.840306 & 85.2  & 1.15   & 149.6 & 47.9  & 59.3  & 0.11   & 11.06  & GV   \\
NGC6338      & 258.845708 & 57.411361 & 118.0 & 1.24   & 16.0  & 66.0  & 150   & 0.25   & 11.56  & GV   \\
NGC6497      & 267.824792 & 59.471000 & 86.2  & 1.16   & 110.3 & 67.1  & 150   & 0.25   & 11.05  & MS   \\
UGC01659     & 32.487042  & 16.032500 & 118.0 & 1.14   & 38.0  & 69.4  & 196   & 0.49   & 10.95  & MS   \\
UGC01938     & 37.092167  & 23.214639 & 90.0  & 1.07   & 155.0 & 81.9  & 186.1 & 0.41   & 10.69  & MS   \\
UGC02134     & 39.715833  & 27.847500 & 64.9  & 1.21   & 106.0 & 71.8  & 150   & 0.39   & 10.88  & MS   \\
UGC02222     & 41.291750  & 32.988417 & 70.1  & 1.12   & 103.0 & 67.6  & 150   & -1.43  & 11.04  & RG   \\
UGC02239     & 41.648500  & 32.449556 & 68.0  & 1.12   & 14.0  & 86.2  & 150   & -0.04  & 10.43  & MS   \\
UGC03960     & 115.094708 & 23.275000 & 31.4  & 1.1    & 46.7  & 72.4  & 150   & -2.57  & 9.90   & RG   \\
UGC04136     & 119.976833 & 47.413306 & 95.3  & 1.2    & 141.5 & 89.9  & 296   & 0.12   & 11.12  & GV   \\
UGC04245     & 122.190750 & 18.194167 & 73.6  & 1.15   & 107.7 & 78.5  & 163   & 0.31   & 10.66  & MS   \\
UGC04262     & 124.764292 & 83.266417 & 81.1  & 1.3    & 151.7 & 40.6  & 188   & 0.20   & 10.84  & MS   \\
UGC04659     & 133.668333 & 47.105139 & 24.2  & 1.15   & 108.5 & 80.0  & 79    & -1.37  & 8.93   & MS   \\
UGC04730     & 135.493292 & 60.151750 & 46.6  & 0.99   & 89.3  & 67.5  & 100   & 0.41   & 10.45  & MS   \\
UGC05326     & 148.852000 & 33.262889 & 18.5  & 0.97   & 146.1 & 16.9  & 60    & -1.05  & 8.79   & MS   \\
UGC05396     & 150.418708 & 10.756389 & 76.7  & 1.17   & 155.9 & 69.9  & 152.6 & -0.07  & 10.53  & MS   \\
UGC08004     & 192.908208 & 31.352944 & 88.3  & 1.04   & 5.6   & 68.7  & 133.9 & -0.01  & 10.30  & MS   \\
UGC08231     & 197.156250 & 54.074611 & 34.8  & 1.17   & 74.9  & 74.4  & 96.8  & -0.39  & 9.44   & MS   \\
UGC08234     & 197.193792 & 62.271694 & 117.0 & 1.21   & 142.1 & 80.2  & 150   & -1.09  & 11.23  & RG   \\
UGC08322     & 198.753875 & 12.725278 & 109.0 & 1.05   & 36.1  & 73.0  & 238.1 & -0.02  & 11.19  & GV   \\
UGC08733     & 207.162458 & 43.412444 & 32.8  & 1.29   & 6.3   & 62.2  & 80.9  & -0.63  & 9.40   & MS   \\
UGC08781     & 208.094708 & 21.539444 & 108.0 & 1.2    & 161.8 & 57.1  & 238.6 & 0.23   & 11.14  & MS   \\
UGC09598     & 223.787833 & 43.818639 & 79.9  & 1.21   & 122.1 & 71.2  & 177   & -0.02  & 10.62  & MS   \\
UGC09629     & 224.296875 & 52.346167 & 112.0 & 1.18   & 153.5 & 76.6  & 100   & -0.57  & 11.16  & RG   \\
UGC10097     & 238.930250 & 47.867306 & 85.2  & 1.15   & 123.9 & 48.2  & 150   & 0.18   & 11.29  & GV   \\
UGC10905     & 263.526625 & 25.344028 & 111.0 & 1.18   & 171.7 & 68.4  & 150   & 0.31   & 11.35  & GV   \\
CGCG163-062  & 217.298625 & 30.077361 & 61.3  & 0.83   & 132.0 & 12.0  & 13.9  & -0.64  & 9.91   & MS   \\
CGCG536-030  & 20.288542  & 40.470722 & 85.1  & 0.87   & 30.5  & 51.3  & 200   & 0.08   & 9.99   & MS   \\
IC0674       & 167.776542 & 43.633056 & 108.0 & 1.14   & 121.0 & 85.3  & 241.9 & 0.03   & 11.08  & GV   \\
IC3598       & 189.337792 & 28.208222 & 110.0 & 1.15   & 139.1 & 76.0  & 225.8 & -0.13  & 11.00  & GV   \\
Mrk1418      & 145.112500 & 48.337528 & 9.5   & 0.82   & 15.8  & 35.6  & 47.1  & -1.30  & 8.52   & MS  \\ 
\end{longtable*}

\subsection{GBT CO\,(1--0) Observations  \label{subsec:observation}} 

For each galaxy in the selected sample, we mapped CO\,(1--0) at a rest frequency of 115.27~GHz across a field of view of $2.5' \times 2.5'$ centered on the galaxy. We used the Argus array receiver \citep{Sieth2014Argus:Telescope} on the GBT with an on-the-fly mapping technique \citep{Mangum2007TheTechnique}. The Argus receiver consists of 16 single-polarization beams, which form a $4\times4$ square array with a side length of $1.52'$ and a spacing of $30.4''$ between beams. This square array also aligns with the elevation and cross-elevation axes of the GBT. 
The 16 beams on Argus are connected to the 16 banks provided by the VEGAS backend \citep{Chennamangalam2014ASpectrometer}.   
Each bank is configured to provide an effective bandwidth of 1.25 GHz with 1024 channels and a spectral resolution of 1.465 MHz, which corresponds to 3.8~km~s$^{-1}$ at the frequency of CO\,(1--0). The angular resolution of our final maps is $8.3''$ (see Section~\ref{subsec:reduction} for further details).

The observations were carried out from November 2021 to March 2025 (project code: GBT21B-024; PI: Bolatto), with a total observing time of 344 hours where 71 galaxies were observed. However, nine of these galaxies\footnote{NGC~0495, NGC~5987, UGC~04054, UGC~04258, UGC~04425, UGC~09777, UGC~8909, UGC~9663, and UGC~9837} were dropped due to poor data quality with severe artifacts\footnote{For data taken in the first year, we identified pointing issues caused by unaccounted movements of GBT's secondary mirror. While this problem was fixed in later observations and we have excluded some of the early sessions, nine of the galaxies are still badly affected and thus are not presented in this paper.}, leaving us with a sample of 62. 
We note that the originally proposed sample includes a total of 150 galaxies, but our observing strategy was to follow up on galaxies that showed hints of detection in order to obtain good maps. Therefore, our final sample of 62 galaxies are possibly biased toward successful detections.   

For each observing session, we started with out-of-focus (OOF) holography scans of a bright source, using either the Ka-band, Q-band, or Argus receiver. The OOF procedure re-aligns the surface panels of the GBT to correct for any residual thermal misalignment, and it typically takes 30--40 minutes. The OOF source was also chosen to be an ALMA flux calibrator\footnote{\url{https://almascience.nrao.edu/sc/}}, so we also used it to take flux calibration scans at our observing frequency after pointing and focusing. Finally, we pointed and focused on a pointing calibration source near our science target to remove residual errors from the telescope pointing model. We then carried out our target observations by alternating mapping scans in the Right Ascension (R.A.) and Declination (Decl.) directions. 

To ensure a consistent pointing and focus throughout our observations, we pointed and focused every 30--40 minutes between each R.A. or Decl.\ mapping scans. For calibration, we employ a vane observation, which uses an ambient temperature load for calibration of the antenna temperature. In Section~\ref{subsec:reduction}, we describe our calibration and imaging processes in detail.

\subsection{GBT CO\,(1--0) Data Reduction  \label{subsec:reduction}} 

The data calibration and gridding are done using an adapted version of \texttt{gbtpipe}\footnote{\url{https://github.com/GBTSpectroscopy/gbtpipe}}, a Python-based pipeline for processing spectral line data and making maps from Argus observations. Our data reduction pipeline code is available in a public GitHub repository, GBT-EDGE\footnote{\url{https://github.com/teuben/GBT-EDGE}}, where the method and procedure for running the pipeline are also documented. We summarize our data reduction process below and refer readers to \citet{2019nrao.reptE...1F} and the GitHub pages for more details. 

First, we retrieve the observed ON, OFF, and vane calibration scans for each galaxy and compute the beam brightness temperature ($T_A^*$) following recommendations in \citet{2019nrao.reptE...1F}:
\begin{equation}
T_A^* = \frac{T_\mathrm{cal}}{C_\mathrm{vane}/C_\mathrm{OFF}(t) - 1} \times
\frac{C_\mathrm{ON}(\nu) - C_\mathrm{OFF}(\nu)}{C_\mathrm{OFF}(\nu)},
\label{eqn_vanecal}
\end{equation}
where $C_\mathrm{ON}$, $C_\mathrm{OFF}$, and $C_\mathrm{vane}$ are instrumental counts from the ON, OFF, and vane calibration (which measures the ambient temperature) scans. 
The left factor represents the time-dependent effective system temperature, which was computed as a scalar for each scan leg, and in the right factor are vector quantities as a function of frequency, computed in this way to improve baseline performance.   

The calibration temperature $T_\mathrm{cal}$ is given by 
\begin{equation}
T_\mathrm{cal} = (T_\mathrm{atm} - T_\mathrm{bg}) + (T_\mathrm{amb} - T_\mathrm{atm})\ \exp(\tau_0 \cdot A),
\label{eqn_Tcal}
\end{equation}
where $T_\mathrm{atm}$ is the atmospheric temperature, $T_\mathrm{bg}$ is the temperature of the cosmic microwave background (2.73~K), $T_\mathrm{amb}$ is the ambient temperature measured from the vane observation, $\tau_0$ is the atmospheric opacity at zenith, and $A$ is the air mass determined
from elevation information. Both the $T_\mathrm{atm}$ and $\tau_0$ are derived from the online GBT weather models\footnote{\url{https://www.gb.nrao.edu/~rmaddale/Weather/index.html}}.   The uncertainty on $T_\mathrm{cal}$ is negligible in comparison to other observational uncertainties, and $T_\mathrm{cal} \approx T_\mathrm{amb}$ within 2\% for the observations.

In observing near $\sim 115$ GHz, the Argus instrument shows variations in bandpass power that change significantly over the course of a pass of the receiver over the target galaxy. Using off-galaxy lines of sight to measure the OFF would require making large maps or dedicating a substantial fraction of the observing time to OFF measurements.  Moreover, with the instability of the bandpass, such observations may not apply to the actual ON component of the spectrum. We thus develop an empirical model for the OFF that can account for time variation over the course of a scan.  

\begin{figure}
\centering
\includegraphics[width=\linewidth]{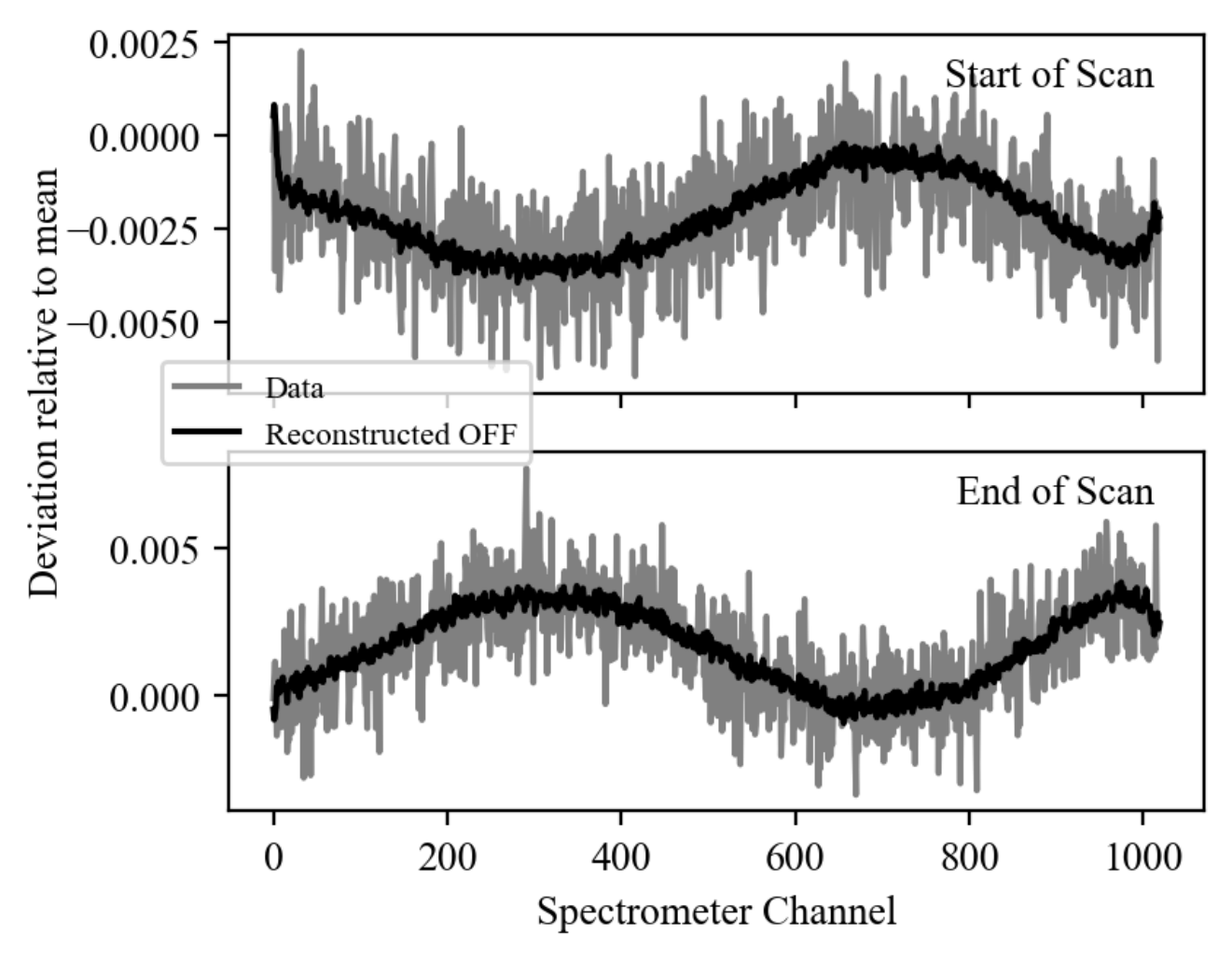}\\
\caption{An example of our background measurements (gray) and empirical OFF model construction (black), represented by the start (top panel) and end (bottom panel) of selected observing scans. These OFF models are used to remove variations in the background signals (Equation~\ref{eqn_Tcal}). This figure exemplifies the changes occurring in the spectral baseline during a leg of the on-the-fly map, caused by a combination of atmospheric and instrument instabilities.}
\label{fig:pca-off}
\end{figure}

We first define a rectangular region on the sky matching the CALIFA field of view\footnote{The CALIFA field of view is a hexagon, and the matched rectangular region is the smallest rectangle that covers that hexagon.} that should contain all galactic emission. For each Argus receiver, we retrieve all position samples in a scan set outside this rectangular zone of exclusion to build an OFF model.  We then apply Principal Component Analysis (PCA) to this set of spectra to find a set of eigenspectra that represent the bandpass variations.  The model retains only those components such that $\sigma^2_{i}/\sigma^2_{i+1}>1.2$ where $\sigma^2_i$ is the variance explained by the $i$th component.  We set the 1.2 threshold empirically to exclude noise-dominated components. 
We typically include 3 to 5 components to represent the OFF.  We then fit all the spectra with these eigenspectra, both inside and outside the zone of exclusion, to generate a model of the OFF and apply Equation \ref{eqn_Tcal}.  Finally, we fit each spectrum with a ninth-order Legendre polynomial baseline to remove residual variations in the spectrum, excluding the velocity range where CO emission is expected.
Figure~\ref{fig:pca-off} demonstrates our PCA-based OFF reconstruction, comparing between the best-fit OFF models (black) and our background OFF data (gray) measured in the start/end of our observing scans. We note that this method allows us to track and remove a time-variable spectral response better than traditional OFF techniques, while the resulting maps are still affected by artifacts likely due to baseline instability at the faint levels (see Section~\ref{subsec:maps} for more details).

After calibration, we produce a spectral line data cube by gridding the data. We first reject bad spectra.  For each spectrum, we determine an empirical noise measurement as $\sigma_T = \langle T_A^*(i) - T_A^*(i+2)\rangle/\sqrt{2}$ where $i$ represents the $i$th channel of the data. Using a two-channel lag avoids the channel-to-channel correlation in the spectrometer. We reject those calibrated spectra that 1) show empirical noise values that are a factor of 1.3 times larger than that expected from the radiometer formula at the measured system temperature, 2) have large-scale baseline ripples with a magnitude 1.3 times larger than the noise, and 3) show any single-channel spikes larger than 5 times the empirical noise.  These thresholds typically reject $\sim 1\%$ of the data. The retained scans are then gridded into a data cube using the Bessel function gridding kernel given in \citet{2007A&A...474..679M}.  Finally, we subtract a seventh-order baseline from each position in the resulting cube, excluding the velocity channels within 400~km~s$^{-1}$ of the galaxy's central velocity. We note that a seventh-order baseline fit is necessary for GBT spectra, as the fitting is done across the entire velocity range of $\sim$3000~km~s$^{-1}$ which is much wider than the galaxy emission scale of $\sim$300~km~s$^{-1}$.  

To achieve a higher signal-to-noise ratio (S/N) without losing the resolution needed to resolve gas distributions, we apply a factor-of-1.3 smoothing in the spatial dimensions as well as a factor-of-4 smoothing in the spectral axis using a Hanning window function. Lastly, the final data cube is converted from the $T_A^*$ scale to the main-beam antenna temperature $T_\mathrm{mb} \simeq T_A^* / \eta_\mathrm{mb}$, where $\eta_\mathrm{mb} = 23\% \pm 3\%$ is the main-beam efficiency.  
We derive $\eta_\mathrm{mb}$ based on measurements of the calibrators 3C84, 3C279, and 3C454.3 during our observing sessions.
First, the aperture efficiency of the GBT is estimated via $\eta_\mathrm{a} = 0.35\ T_\mathrm{A}^* / S_\nu$, where $T_\mathrm{A}^*$ is measured from the amplitude of peak scans and $S_\nu$ is the flux density of the calibrator source, which we obtain from the ALMA Calibration Catalog. We derive a mean $\eta_\mathrm{a}$ of $12.5\% \pm 2.2\%$, which corresponds to a surface rms error of 280~$\mathrm{\mu m}$ based on the Ruze equation \citep{1966ruze}.
Then, with $\eta_\mathrm{a}$ and assuming a Gaussian beam, $\eta_\mathrm{mb}$ can be derived as
\begin{equation}
\eta_\mathrm{mb} = 0.89\ \eta_\mathrm{a}\ \left( \theta_\mathrm{FWHM} \frac{D}{\lambda}\right)^2 ,
\label{eqn_etamb}
\end{equation}
where $\theta_\mathrm{FWHM}$ is the beam size in radians measured from the FWHM of peak scans, the GBT dish size $D$ is 100~m, and the wavelength $\lambda$ is 2.6~mm for the CO\,(1--0) line.
More details on these equations and derivations can be found in \citet{2019nrao.reptE...1F}.

Our final data cubes have a spectral resolution of 15.2~km~s$^{-1}$ and a spatial resolution of $8.3''$, which corresponds to $\sim$2 kpc for a typical distance of 60--70 Mpc across our sample. The $8.3''$ beam size can be derived via $1.18  \times 1.3 \times \lambda / D$, where 1.18 comes from the GBT optics, and 1.3 is the post-gridding smoothing.
We note that due to the uneven integration coverage inherent to the Argus beam layout, the outer regions of our mapping area naturally receive less integration time and therefore have higher noise. Thus, in Section~\ref{sec:result}, our map products are produced with S/N-based masking, and our results only focus on inner mapping areas around the galaxies.  

\begin{figure*}
\centering
\includegraphics[width=\linewidth]{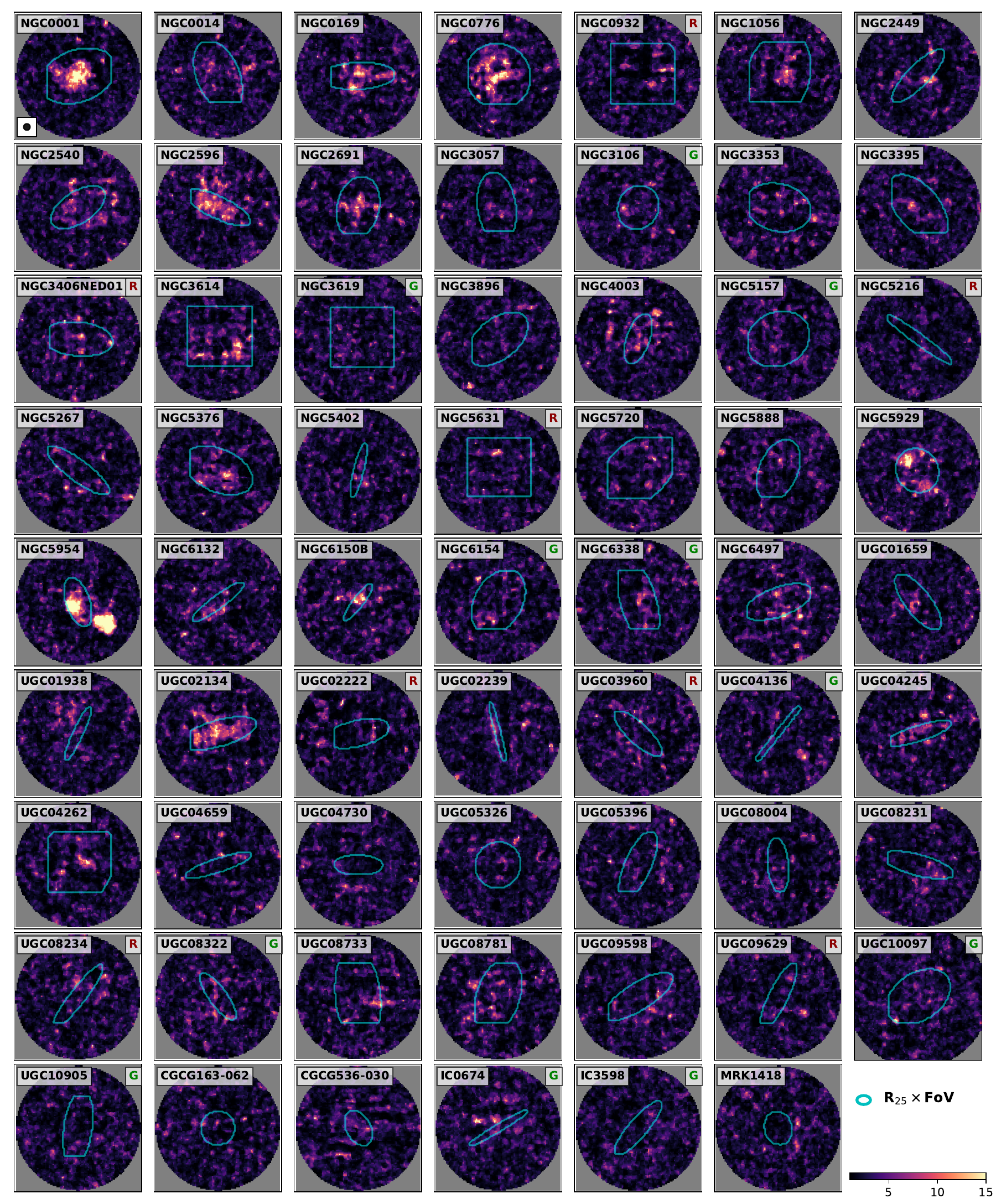}\\
\caption{The peak signal to noise (S/N) maps of all \Ngal galaxies, integrating over a channel width of 90 km~s$^{-1}$ to highlight gas structures with line widths at that scale. GV and RGs are labeled by `G' and `R', respectively, on the top-right corner. The color scale (bottom-right panel) ranges from S/N = 1--15. Each panel shows the entire GBT field of view of $2.5' \times 2.5'$, and the common beam size of $8.3''$ ($\sim$2 kpc for our sample) is shown on the top-left panel. The overlaid cyan contours indicate the intersection between each galaxy's R$_{25}$ radius and CALIFA's field of view, which defines the area for computing galaxy-integrated quantities (Section~\ref{sec:result}).}
\label{fig:tpeak}
\end{figure*}

Figure~\ref{fig:tpeak} shows a compilation of all the peak S/N maps derived from the final data cubes. In these maps, we smooth the channel widths to be 6 times wider ($\sim$90 km~s$^{-1}$) so that they better reflect molecular gas features with line widths on a similar scale. The cross section of each galaxy's projected R$_{25}$ radius (derived from d$_{25}$, inclination, and P.A. in Table~\ref{tab:sample}) and the CALIFA field of view is overlaid as cyan contours, which define the areas for our analysis on galaxy-integrated quantities (e.g., star formation rates, molecular gas masses, and depletion times) in Section~\ref{sec:result}.

\begin{figure}
\centering
\includegraphics[width=.9\linewidth]{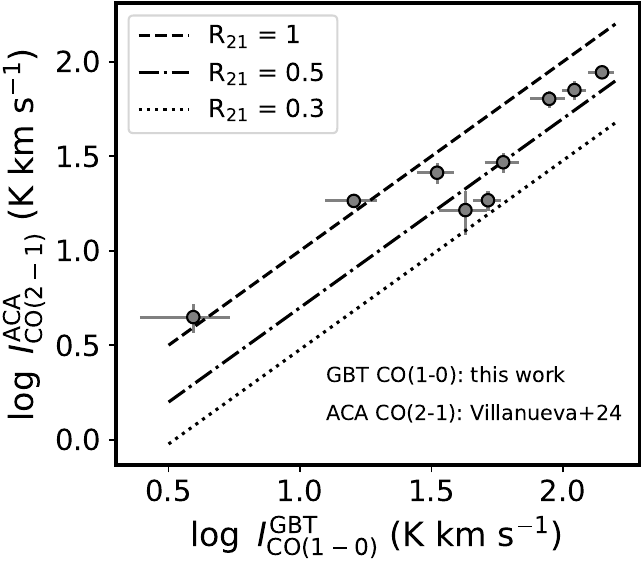}\\
\caption{Comparison of the GBT CO\,(1--0) integrated line fluxes with those from the ALMA ACA CO\,(2--1) data \citep{2024ApJ...962...88V} for the nine galaxies that overlap between both samples. They show a CO (2--1)/(1--0) line ratio ($R_{21}$) of $0.70 \pm 0.27$. Galaxies with the highest and lowest $R_{21}$ tend to have higher uncertainties in the GBT data.}
\label{fig:flux-compare}
\end{figure}

As a check on our data reduction procedure, we compare the galaxy-integrated CO line fluxes from the GBT to those from \citet{2024ApJ...962...88V}, where nine of their galaxies overlapped with our sample and were observed in CO\,(2--1) with the ALMA ACA array at $12''$ resolution. 
These nine galaxies are NGC~0001, NGC~0169, NGC~1056, NGC~2540, NGC~2596, UGC~04245, UGC~05396, UGC~08322, and UGC~08781, all of which are MS galaxies except for UGC~08322 (see Table~\ref{tab:sample}).
For a consistent comparison, we also convolve the GBT data of these galaxies to $12''$.
As shown in Figure~\ref{fig:flux-compare}, the CO\,(2--1)/(1--0) line ratios ($R_{21}$) of these galaxies generally fall between 0.5--1.0, having a median and standard deviation of 0.70 and 0.27, respectively. This is consistent with the $R_{21}$ values observed across nearby MS galaxies \citep{2021MNRAS.504.3221D,2021PASJ...73..257Y,2022ApJ...927..149L}.

\subsection{Optical IFU Data from CALIFA}

Our analyses also make use of various properties measured by the CALIFA survey \citep{2016A&A...594A..36S,2023MNRAS.526.5555S}, which obtained optical IFU spectroscopy with the Potsdam Multi Aperture Spectograph/PPak instrument mounted on the 3.5-m telescope at the Calar Alto Observatory. 
These optical IFU data have an effective spatial resolution of $\sim$1.5$''$, a field of view of $74'' \times 64''$, and uncertainties of $\sim$9\% in the absolute photometric calibration and $\sim$4\% in the blue-to-red spectro-photometric precision \citep{2023MNRAS.526.5555S}.

We use the data products generated by \texttt{Pipe3D} \citep{2016RMxAA..52..171S,2016RMxAA..52...21S}, which is a widely used pipeline for analyzing the spectroscopic properties of stellar populations and ionized gas measured from optical IFU surveys. 
For our galaxy sample, we adopt the latest \texttt{Pipe3D} data products from the extended data release \citep[eDR;][]{2023MNRAS.526.5555S}. However, an exception is NGC~2596, which is excluded in eDR due to technical issues in its image reconstruction. Therefore, we use an earlier data product release from DR3 \citep{2016A&A...594A..36S} only for this galaxy.

From these \texttt{Pipe3D} data products we obtain multiple emission line-related properties used in this work, including their fluxes, velocities, equivalent widths, and associated errors. We also obtain resolved dust-corrected stellar masses ($M_\mathrm{star}$) and star formation histories (SFH), which provide luminosity fractions across various bins of age and metallicity from the stellar population. We refer readers to Section~\ref{subsec:estimation} and \citet{2016RMxAA..52..171S} for more details on these quantities.

\section{Results  \label{sec:result}} 

In this section, we present the procedure and results of our map products and scientific analyses. We publish all the reduced GBT data cubes and maps online\footnote{\url{https://doi.org/10.5281/zenodo.20707368}}. We also release all of our calculations and analysis code in a public GitHub repository\footnote{\url{https://github.com/ElthaTeng/gbt-edge-analysis} and Zenodo: \url{https://doi.org/10.5281/zenodo.20707910} \label{gbt-edge-analysis}}.

\subsection{CO Moment Maps and Error Estimation \label{subsec:maps}} 

From the GBT CO\,(1--0) data cubes, we derive moment maps for all \Ngal observed galaxies. These maps include the integrated intensity (moment-0), intensity-weighted velocity (moment-1), and the velocity dispersion (moment-2). 
For almost all of our sample (except for Mrk1418), we use two different methods to create signal masks. These signal masks are used to select the pixels and channels within the data cubes for moment map productions. We then include pixels and channels covered by either or both (i.e., the logical ``inclusive OR'') of these signal masks for creating the final moment maps. 
 
First, we create the `H$\alpha$ masks'. Assuming that CO and H$\alpha$ have similar kinematics \citep{2018ApJ...860...92L,2022ApJ...934..173S}, we construct a signal mask based on the H$\alpha$ velocity field obtained from the extended CALIFA data release \citep{2023MNRAS.526.5555S}. Based on the CALIFA \texttt{Pipe3D} data, we select regions with well-detected central H$\alpha$ velocity ($v_0$) and include a spectral range of $v_0\pm$FWHM, where FWHM follows a radial-dependent relation found across a sub-sample of CALIFA galaxies based on CO\,(1--0) data from CARMA \citep{2021ApJ...923...60V}. Finally, we re-project the mask onto the GBT data grid, using nearest-neighbor matching implemented in the Astropy-affiliated \texttt{reproject} package. This masking approach is similar to the H$\alpha$ mask adopted in \citet{2024ApJ...962...88V}, and it has the advantage of providing a robust kinematic guidance.

Next, we create the `CO-dilated masks'.
Based on the reduced CO data cubes, we initially include all position-position-velocity (ppv) spaces with emission above $2.5\sigma$\footnote{The CO-dilated mask starting with $2.5\sigma$ is found to result in most consistent fluxes between the GBT and ALMA ACA data, as shown in Section~\ref{subsec:reduction}}. Then, we expand the mask coverage by twice the beam size in the spatial directions and $\pm 30$~km~s$^{-1}$ in the spectral direction (i.e., two adjacent channels). We also exclude spectral regions outside $\pm 2\times$\texttt{vmaxg}\footnote{reported in HyperLeda: \url{http://leda.univ-lyon1.fr/} \label{hyperleda}}, which ensures covering only the spectral range within the observed maximum rotation velocity of atomic gas. For 12 of the galaxies in our sample without a measured \texttt{vmaxg}, we assume a conservative \texttt{vmaxg} value of 150~km~s$^{-1}$ that is slightly larger than the MW value where $2\times$\texttt{vmaxg} $\sim 250$~km~s$^{-1}$ \citep{2016ARA&A..54..529B,2016ApJ...832..159R}. This CO-dilated masking procedure is very similar to that adopted in \citet[][see also \citealt{2006PASP..118..590R,2024ApJS..271...35W}]{2017ApJ...846..159B}. 

By adopting both the H$\alpha$ and CO-dilated masks for moment map creation, our selections of pixels and channels avoid biases from only the H$\alpha$ or CO data.
For Mrk~1418, however, we use an alternative signal mask (i.e., the `block mask') due to poor S/N in its H$\alpha$ and CO data.
To obtain the block mask, we simply select all pixels within the R$_{25}$ radius and include all channels within $\pm2\times$\texttt{vmaxg}. While the block mask is model-independent, it results in high noise because of its broad coverage.

\begin{figure*}
\begin{minipage}{\linewidth}
\centering
\includegraphics[width=\linewidth]{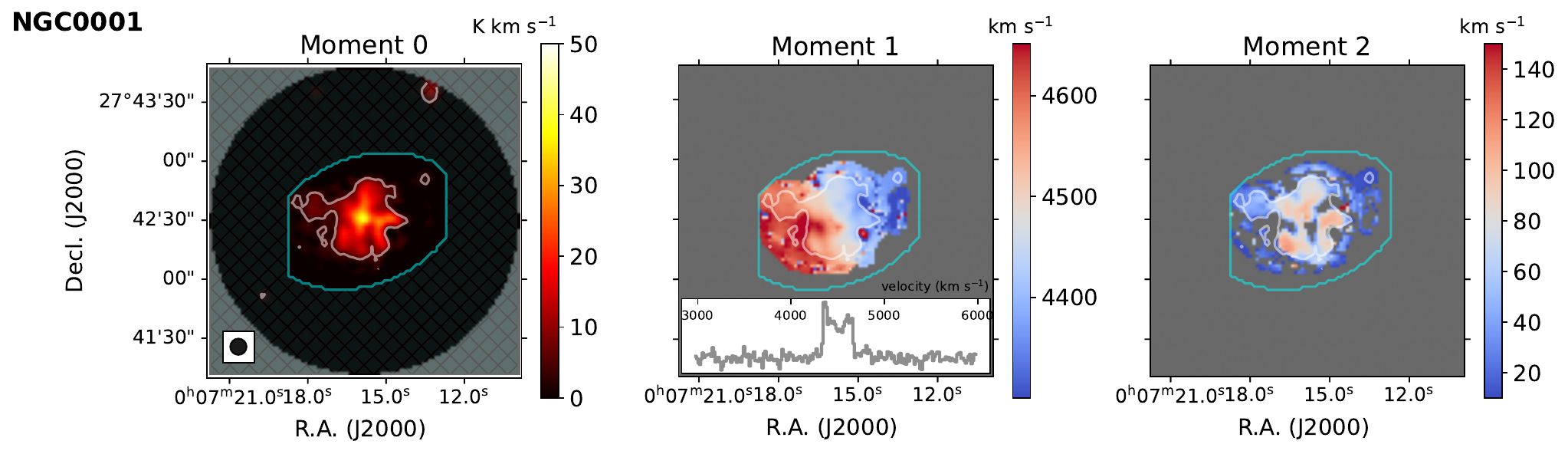}
\end{minipage}\\
\begin{minipage}{\linewidth}
\centering
\includegraphics[width=\linewidth]{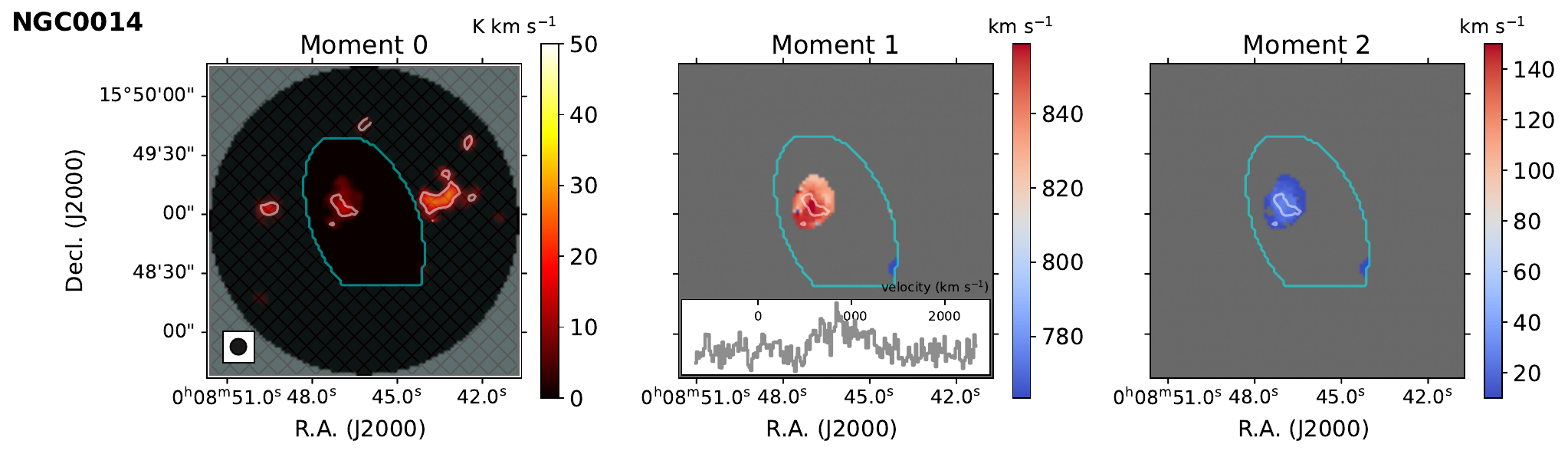}
\end{minipage}
\caption{Examples of the moment 0 (left), 1 (middle), and 2 (right) maps, based on H$\alpha$ + CO-dilated masking (see Section~\ref{subsec:maps}). The white contours show S/N $> 5$ based on the moment-0 maps. The common beam size and cyan contours are the same as in Figure~\ref{fig:tpeak}. Some galaxies show high-quality detection in moment 0 and clear velocity gradients indicative of galaxy rotation (e.g., top row, NGC~0001), while some galaxies show spurious emission outside their $R_{25}$ radii due to baseline variations, leading to high flux uncertainties (e.g., bottom row, NGC~0014). Our analysis only includes regions within the $R_{25}$ radii, as enclosed by the cyan contours. The integrated spectra of S/N $> 5$ regions within the cyan contours are inserted into the middle panel.}
\label{fig:mom0-datapref}
\end{figure*}

We apply all the above-mentioned signal masks and create corresponding moment 0, 1, and 2 maps using the \texttt{spectral-cube} package.
Figure~\ref{fig:mom0-datapref} presents the moment maps for NGC~0001 and NGC~0014 as a demonstration of the overall data quality. 
While we have obtained high-quality data with reliably detected signals in some galaxies (e.g., NGC~0001), we also note that a few galaxies show somewhat dubious emission outside the galaxy's $R_{25}$ (e.g., NGC~0014)\footnote{excluding the cases of NGC~0169, 5929, and 5954 since they are interacting galaxies}.
Inspection of the spectra in those regions shows that most of the emission outside $R_{25}$ is spurious, likely due to remaining baseline variations across the data cubes that falsely boost S/Ns at certain ppv locations. We do not include those regions in the analysis, and we account for similarly spurious emission within $R_{25}$ (i.e., our regions of interest) in our uncertainty estimates, as described in the following paragraphs. The galaxies with spurious emission tend to have higher flux uncertainties, which can make estimated total fluxes become upper limits (e.g., see NGC~0014 in Table~\ref{tab:derived}).

The moment maps for all GBT-EDGE galaxies are presented in Appendix~\ref{appendix:maps} as a figure set. Overall, the moment-0 maps reveal extended molecular gas structures across all galaxies, strengthening many gas features seen in Figure~\ref{fig:tpeak}. In addition, these moment-0 maps show that CO emission concentrates not only toward galaxy centers, but a significant amount of gas is also found throughout the outer disks.
The moment-1 maps reveal clear signatures of rotation across many of the spiral or disk galaxies. The moment-2 maps show CO velocity dispersion up to 100--150~km~s$^{-1}$ across the sample, which is expected for molecular gas observations at $>$~kpc scales \citep[e.g.,][]{2017ApJ...846..159B,2020ApJ...903..145L}.

For all the maps, we also produce their respective uncertainty maps. First, we estimate the rms noise ($\sigma_\mathrm{rms}$) for each pixel by calculating the standard deviation of all the signal-free channels. Here we approximate signal-free channels as those with nothing above three times the rms value across the entire spectrum. Then, the integrated uncertainty $\sigma_\mathrm{int}$ per pixel is calculated via Gaussian error propagation: $\sigma_\mathrm{int} = \sigma_\mathrm{rms} \cdot \Delta v \cdot \sqrt{N}$, where $\Delta v$ is the channel width ($\sim$15 km~s$^{-1}$) and $N$ is the number of channels integrated with different masking schemes. 
Summing $\sigma_\mathrm{int}$ over the $R_{25}$ of each galaxy, the integrated flux uncertainty ($\sigma_\mathrm{flux}$) is typically $\sim$20\% of the total flux, with a 16$^\mathrm{th}$--84$^\mathrm{th}$ percentile range of 16--29\% across our sample. 
This $\sigma_\mathrm{flux}$ is an estimate of the formal error, but it does not include the effect of baseline and mask production systematics.

In addition to $\sigma_\mathrm{flux}$, we also evaluate the uncertainties induced by  our masking procedure as well as baseline variations across our data cubes, $\sigma_\mathrm{add}$. We assess $\sigma_\mathrm{add}$ by comparing measured fluxes after a `fake source' is added to a series of random locations in our data cubes.
For each galaxy, we start by selecting a random ppv position within the defined `block' mask. At that position, we insert a simulated Gaussian emission structure with a peak intensity of 1~K, a size of the beam size, and a typical line FWHM of 60~km~s$^{-1}$. 
We adopt this set of characteristics for the fake source because it is representative of the spurious emission seen outside $R_{25}$.  
Next, we run our H$\alpha$ + CO-dilated masking routine and measure the total flux including both the original and artificial sources. 
Then, we subtract the originally measured flux (without the inserted source) from the measured total flux. The above procedure is repeated for 100 times per galaxy.
Finally, we compare the average measured flux to the known flux of the inserted source to estimate $\sigma_\mathrm{add}$ in terms of percentage for each galaxy.
We obtain a median $\sigma_\mathrm{add}$ of 51\% across all \Ngal galaxies, while the 16$^\mathrm{th}$--84$^\mathrm{th}$ percentile values span from 16--144\% depending on the data quality of individual galaxy. 

To obtain our final error estimation, we add the additional uncertainty $\sigma_\mathrm{add}$ quadratically to the previously derived $\sigma_\mathrm{flux}$, namely, $\sigma_\mathrm{final} = \sqrt{\sigma_\mathrm{flux}^2 + \sigma_\mathrm{add}^2}$.
For 15 of the galaxies in our sample, their final uncertainties exceed 100\% of the measured fluxes, and thus their inferred CO luminosities are presented as upper limits (see Table~\ref{tab:derived}). 
We note that $\sigma_\mathrm{add}$ and $\sigma_\mathrm{flux}$ are not independent, as our $\sigma_\mathrm{add}$ estimates likely contain contributions from $\sigma_\mathrm{flux}$. This means that our final uncertainty $\sigma_\mathrm{final}$ could be overestimated and be overly conservative.

\subsection{Derivation of Fundamental Quantities   \label{subsec:estimation}}

To study the molecular gas depletion time ($t_\mathrm{dep} \equiv M_\mathrm{mol}$ / SFR) or star formation efficiency (SFE $= 1 / t_\mathrm{dep}$) across our galaxy sample, we derive the star formation rates (SFR) and molecular gas masses ($M_\mathrm{mol}$) for all galaxies. In this subsection, we describe the methods and prescriptions adopted to estimate these quantities. We then compare the resulting $t_\mathrm{dep}$ and discuss their implications in Section~\ref{subsec:sfe}.

\subsubsection{Star Formation Rates  \label{subsubsec:sfr}}  

Based on the Balmer decrement and assuming the extinction curve from \citet{1989ApJ...345..245C}, the extinction of H$\alpha$ can be estimated via 
\begin{equation}
A_\mathrm{H\alpha} = 5.86\ \log\left( \frac{F_\mathrm{H\alpha}}{2.86\,F_\mathrm{H\beta}} \right) ,
\label{eqn_extinction}
\end{equation}
where $F_\mathrm{H\alpha}$ and $F_\mathrm{H\beta}$ are the corresponding Balmer line fluxes, and 2.86 represents their nominal flux ratio for case-B recombination. The extinction-corrected SFRs in units of M$_\odot$~yr$^{-1}$ can then be computed by 
\begin{equation}
\mathrm{SFR} = 7.9 \times 10^{-42} \cdot F_\mathrm{H\alpha} \cdot 10^{A_\mathrm{H\alpha}/ 2.5} ,
\label{eqn_sfr_ha}
\end{equation}
where a \citet{1955ApJ...121..161S} initial mass function (IMF) is assumed \citep[and following e.g., \citealt{2020A&A...644A..97C,2023MNRAS.526.5555S}]{2002MNRAS.332..283R}.

With Equations~\ref{eqn_extinction} and~\ref{eqn_sfr_ha}, we use the measured Balmer line properties from the CALIFA \texttt{Pipe3D} products and derive resolved H$\alpha$-based SFR maps for our galaxies. We exclude non-star-forming pixels where the H$\alpha$ equivalent width is less than 6~$\mathrm{\AA}$ \citep[e.g.,][]{2020MNRAS.492.3073L,2021MNRAS.503.1615S}. To avoid unrealistic SFR values from low S/N measurements, we also set $A_\mathrm{H\alpha} = 3$~mag for all pixels with derived $A_\mathrm{H\alpha} > 3$~mag \citep[e.g.,][]{2017ApJ...846..159B}. Finally, to obtain an integrated SFR value for each galaxy, we co-add the resolved SFRs over all pixels within the R$_{25}$ radius. We also error propagate the uncertainties and obtain a $< 0.1$~dex error for these global SFR values. Such level of error is also consistent with the CALIFA global table in \citet{2024ApJS..271...35W}.  

Since our galaxy sample includes a significant fraction of quenched galaxies with low levels of star formation, the H$\alpha$-inferred SFRs in these systems could be overestimated due to possible ionization from old stars, AGNs, or shocks \citep[e.g., review by][]{2012ARA&A..50..531K}. Therefore, in addition to H$\alpha$, we also use a simple stellar population (SSP) analysis to derive SFRs based on the young stellar population. 
In Appendix~\ref{appendix:ssp}, we show that SFRs estimated from both methods agree well for MS and GV galaxies. For RGs, however, we find that H$\alpha$-based SFRs provide significantly better constraints (i.e., stricter upper limits in SFRs) than using the SSP analysis.

In summary, while we find that SSP does not provide improved constraints on SFRs across our RG sample, the consistency seen across the MS and GV sample suggests that H$\alpha$ as a SFR indicator can be extended to GV galaxies and may still be useful for RGs compared to using SSP. Thus, we adopt H$\alpha$-based SFRs in all our following analyses and results. We caveat that the H$\alpha$-based SFRs are likely still overestimated for quenched galaxies, and we will discuss in Section~\ref{subsec:uncertainty} how such uncertainties may impact our results and interpretations.

\subsubsection{Molecular Gas Masses  \label{subsubsec:Mgas}} 

The common approach to measure molecular gas in galaxies is to convert observed CO\,(1--0) line fluxes to the total H$_2$ mass via assuming a CO-to-H$_2$ conversion factor (\aco). The value of \aco\ in the Milky Way (MW) disk has been found to be $\approx$ 4.35~$\mathrm{M_\odot\ (K~km~s^{-1}~pc^2)^{-1}}$ \citep[see review by][]{co-to-h2}, and thus many past studies simply adopt this constant value to obtain the amount of molecular gas in galaxies. However, \aco\ does vary within and between galaxies \citep{2013ApJ...777....5S,2022ApJ...925...72T,2023ApJ...950..119T,2023A&A...676A..93D,2025AJ....169...18D,2023PASJ...75..743Y,2024ApJ...964...18C}, and its value depends heavily on local gas conditions such as metallicity, temperature, density, and dynamical properties \citep{2012MNRAS.421.3127N,2012ApJ...751...10P,co-to-h2,2019A&A...621A.104R,2020ApJ...903..142G,2023ApJ...950..119T}. 

To account for the variations in \aco, recent studies have proposed various \aco\ prescriptions which include the environmental dependence of \aco\ and can predict \aco\ based on observable quantities such as metallicity, stellar mass surface density, and CO-line related properties \citep{co-to-h2,2017MNRAS.470.4750A,2020ApJ...903..142G,2024ARA&A..62..369S,2024ApJ...961...42T}.
These prescriptions can be applied to large galaxy samples, which have revealed systematic effects on SFE or other star formation-related quantities across galaxies that were previously unknown due to adopting a constant \aco\ \citep{2020MNRAS.492.6027E,2023A&A...680A...4Q,2023ApJ...945L..19S,2024ApJ...961...42T}. 

In this work, we derive molecular gas masses ($M_\mathrm{mol}$) using four different \aco\ treatments, aiming to reduce possible biases caused by \aco\ variations in our molecular gas measurements. In other words, we will obtain four estimates of $M_\mathrm{mol}$ using four versions of \aco\ based on the following equation: 
\begin{equation}
M_\mathrm{mol} = \alpha_\mathrm{CO} \cdot L'_\mathrm{CO(1-0)} = \alpha_\mathrm{CO} \cdot I_\mathrm{CO(1-0)} \cdot A_\mathrm{pc^2}~, 
\label{def_alphaCO}
\end{equation}
where $L'_\mathrm{CO(1-0)}$ is the line luminosity of CO $J$=1--0 (in units of K~km~s$^{-1}$~pc$^2$) and can be determined by integrating our moment-0 maps ($I_\mathrm{CO(1-0)}$) over a surface area in pc$^2$ ($A_\mathrm{pc^2}$, which depends on the target's distance, see Table~\ref{tab:sample} and Equation~3 in \citealt{co-to-h2}).
The derived $L'_\mathrm{CO(1-0)}$ and \aco~values are listed in Table~\ref{tab:derived}, which together provide our $M_\mathrm{mol}$ estimates via Equation~\ref{def_alphaCO}. The associated uncertainty for each galaxy is error propagated from the final flux uncertainty ($\sigma_\mathrm{final}$) as described in Section~\ref{subsec:maps}.
The statistics of \aco~and $M_\mathrm{mol}$ among MS, GV, and RG groups are presented in Table~\ref{tab:statistics}. We present the details for our \aco\ implementations in Section~\ref{subsubsec:aco}.

\setlength{\tabcolsep}{1pt}

\begin{longtable*}[c]{lcccccccccc}
\caption{Derived Global Properties of the GBT-EDGE Galaxies \label{tab:derived}} \\
\hline \hline
\multirow{2}{*}{Name}  & \multirow{2}{*}{$a$}  & \multirow{2}{*}{$b$}  & $\log L'_\mathrm{CO(1-0)}$ & \multicolumn{3}{c}{\aco~[$\frac{\mathrm{M}_\odot}{\mathrm{K~km~s}^{-1}~\mathrm{pc}^2}$]}   & \multicolumn{4}{c}{$t_\mathrm{dep}$ [Gyr]}  \\ 
\cmidrule(lr){5-7} \cmidrule(lr){8-11}
&   &   & [K~km~s$^{-1}$~pc$^2$] & B13 & SL24 & T24* & MW & B13 & SL24 & T24* \\
(1) & (2)  & (3)  & (4) & (5) & (6) & (7) & (8) & (9) & (10) & (11) \\
\hline
\endfirsthead
\endhead

\toprule
\multirow{2}{*}{Name}  & \multirow{2}{*}{$a$}  & \multirow{2}{*}{$b$}  & $\log L'_\mathrm{CO(1-0)}$ & \multicolumn{3}{c}{\aco~[$\frac{\mathrm{M}_\odot}{\mathrm{K~km~s}^{-1}~\mathrm{pc}^2}$]}   & \multicolumn{4}{c}{$t_\mathrm{dep}$ [Gyr]}  \\ 
\cmidrule(lr){5-7} \cmidrule(lr){8-11}
&   &   & [K~km~s$^{-1}$~pc$^2$] & B13 & SL24 & T24* & MW & B13 & SL24 & T24* \\
(1) & (2)  & (3)  & (4) & (5) & (6) & (7) & (8) & (9) & (10) & (11) \\
\midrule
\endhead
\midrule
\multicolumn{11}{r}{{Continued on next page}} \\
\endfoot

\bottomrule \\
\multicolumn{11}{l}{\parbox{\dimexpr\textwidth-2\tabcolsep}{%
     \textbf{Note.} (1) Galaxy name; (2, 3) best-fit coefficients for metallicity gradient (Equation~\ref{eqn_Zprime_fit}); (4) integrated CO luminosities with errors propagated from $\sigma_\mathrm{final}$ as described in Section~\ref{subsec:maps}; all the upper limits have $> 0.3$~dex uncertainty; (5--7) integrated \aco~values using B13, SL24, and T24* prescriptions (Equations~\ref{eqn_alpha_B13}--\ref{eqn_aco_T24}), with expected \aco~uncertainties of 0.2--0.3~dex; (8--11) global gas depletion time based on \aco~assumptions of MW (\aco\ $= 4.35$ M$_\odot$~(K~km~s$^{-1}$~pc$^2$)$^{-1}$), B13, SL24, and T24*, respectively. The errors in $t_\mathrm{dep}$ are propagated from the flux error ($\sigma_\mathrm{flux}$) in Section~\ref{subsec:maps}.}} \\
\endlastfoot

NGC0001       & -0.00397   & 0.101         & 9.26 $\pm$ 0.17                 & 3.77       & 3.64        & 1.95        & 2.10 $\pm$ 0.28  & 1.82 $\pm$ 0.25  & 1.76 $\pm$ 0.25  & 0.94 $\pm$ 0.15 \\
NGC0014       & 0.0017   & -0.146         & $<$ 6.88                  & 4.85       & 6.37        & 5.94        & 0.95 $\pm$ 0.25  & 1.05 $\pm$ 0.28  & 1.39 $\pm$ 0.37  & 1.29 $\pm$ 0.36 \\   
NGC0169       & -0.00286   & 0.0711         & 9.59 $\pm$ 0.14                 & 3.65       & 5.10        & 1.61        & 8.82 $\pm$ 1.04  & 7.39 $\pm$ 0.96  & 10.30 $\pm$ 1.29  & 3.25 $\pm$ 0.43 \\
NGC0776       & -0.00254   & 0.0831         & 9.24 $\pm$ 0.09                 & 3.85       & 3.77        & 2.26        & 2.56 $\pm$ 0.46  & 2.27 $\pm$ 0.40  & 2.22 $\pm$ 0.39  & 1.33 $\pm$ 0.27 \\
NGC0932       & -0.000521   & -0.0331         & 8.84 $\pm$ 0.28                 & 4.00       & 4.56        & 1.88        & 7.05 $\pm$ 1.31  & 6.48 $\pm$ 1.19  & 7.38 $\pm$ 1.35  & 3.04 $\pm$ 0.64 \\
NGC1056       & -0.00219   & -0.00685         & $<$ 8.18                 & 4.01       & 4.79        & 2.06        & 0.61 $\pm$ 0.10  & 0.56 $\pm$ 0.10  & 0.67 $\pm$ 0.12  & 0.29 $\pm$ 0.05 \\  % 8.81 $\pm$ 0.51
NGC2449       & -0.0036   & 0.0732         & 9.01 $\pm$ 0.09                 & 4.17       & 4.17        & 1.94        & 6.56 $\pm$ 1.09  & 6.28 $\pm$ 1.03  & 6.30 $\pm$ 1.04  & 2.92 $\pm$ 0.57 \\
NGC2540       & -0.00429   & 0.0726         & 9.20 $\pm$ 0.13                 & 4.31       & 4.58        & 1.89        & 2.45 $\pm$ 0.47  & 2.43 $\pm$ 0.46  & 2.58 $\pm$ 0.49  & 1.07 $\pm$ 0.25 \\
NGC2596       & -0.00778   & 0.185         & $<$ 9.19                 & 3.04       & 3.52        & 4.02        & 1.17 $\pm$ 0.19  & 0.82 $\pm$ 0.13  & 0.95 $\pm$ 0.16  & 1.08 $\pm$ 0.16 \\  % 9.83 $\pm$ 0.30
NGC2691       & -0.00297   & 0.0504         & 8.93 $\pm$ 0.21                 & 4.06       & 4.18        & 1.40        & 2.46 $\pm$ 0.39  & 2.29 $\pm$ 0.37  & 2.36 $\pm$ 0.38  & 0.79 $\pm$ 0.14 \\
NGC3057       & 0.000182   & -0.165         & $<$ 7.66                 & 5.17       & 7.58        & 3.43        & 1.35 $\pm$ 0.32  & 1.60 $\pm$ 0.39  & 2.34 $\pm$ 0.56  & 1.06 $\pm$ 0.31 \\  
NGC3106       & -0.000901   & -0.017         & 8.84 $\pm$ 0.26                 & 4.29       & 4.83        & 2.54        & 4.41 $\pm$ 1.20  & 4.36 $\pm$ 1.19  & 4.89 $\pm$ 1.33  & 2.57 $\pm$ 0.73 \\
NGC3353       & 0.00469   & -0.307         & $<$ 7.03                & 5.36       & 9.23        & 4.97        & 0.10 $\pm$ 0.03  & 0.12 $\pm$ 0.03  & 0.21 $\pm$ 0.06  & 0.11 $\pm$ 0.03 \\   
NGC3395       & -0.000502   & -0.0821         & $<$ 7.98                 & 4.58       & 5.84        & 2.38        & 0.63 $\pm$ 0.12  & 0.66 $\pm$ 0.13  & 0.84 $\pm$ 0.16  & 0.34 $\pm$ 0.08 \\  
NGC3406NED01  & -0.00144   & 0.00597         & $<$ 9.10                 & 3.95       & 4.29        & 3.17        & 84.30 $\pm$ 16.80  & 76.40 $\pm$ 15.30  & 82.90 $\pm$ 16.70  & 61.20 $\pm$ 14.20 \\  
NGC3614       & -0.00161   & -0.000114         & $<$ 8.74                 & 3.76       & 4.36        & 1.48        & 4.32 $\pm$ 0.67  & 3.74 $\pm$ 0.55  & 4.32 $\pm$ 0.63  & 1.47 $\pm$ 0.31 \\  
NGC3619       & 0.0000423   & -0.0353         & $<$ 7.38                 & 2.95       & 3.26        & 3.91        & 1.91 $\pm$ 0.50  & 1.30 $\pm$ 0.32  & 1.43 $\pm$ 0.35  & 1.72 $\pm$ 0.48 \\  
NGC3896       & -0.000233   & -0.0891         & $<$ 6.68                 & 4.70       & 6.03        & 5.82        & 1.13 $\pm$ 0.31  & 1.22 $\pm$ 0.33  & 1.57 $\pm$ 0.43  & 1.52 $\pm$ 0.44 \\  
NGC4003       & -0.00155   & -0.00594         & 9.09 $\pm$ 0.19                & 4.15       & 4.68        & 1.59        & 6.06 $\pm$ 1.13  & 5.79 $\pm$ 1.09  & 6.52 $\pm$ 1.23  & 2.22 $\pm$ 0.45 \\  
NGC5157       & -0.00149   & 0.0212         & 8.82 $\pm$ 0.16                 & 4.08       & 4.44        & 2.76        & 5.00 $\pm$ 1.79  & 4.68 $\pm$ 1.69  & 5.10 $\pm$ 1.83  & 3.17 $\pm$ 1.02 \\
NGC5216       & 0.000111   & -0.0629         & 7.76 $\pm$ 0.09                 & 4.55       & 5.34        & 5.13        & 171.0 $\pm$ 34.4  & 180.0 $\pm$ 35.9  & 210.0 $\pm$ 42.1  & 202.0 $\pm$ 52.4 \\
NGC5267       & -0.00158   & 0.0522         & 8.81 $\pm$ 0.15                 & 4.07       & 4.00        & 4.35        & 3.52 $\pm$ 1.02  & 3.29 $\pm$ 0.97  & 3.23 $\pm$ 0.95  & 3.52 $\pm$ 1.00 \\
NGC5376       & -0.00287   & 0.103         & 8.56 $\pm$ 0.23                 & 3.71       & 3.52        & 2.03        & 2.40 $\pm$ 0.39  & 2.05 $\pm$ 0.34  & 1.94 $\pm$ 0.32  & 1.12 $\pm$ 0.21 \\
NGC5402       & -0.00212   & -0.00142         & 8.02 $\pm$ 0.19                 & 4.08       & 4.58        & 3.56        & 0.41 $\pm$ 0.10  & 0.38 $\pm$ 0.10  & 0.43 $\pm$ 0.11  & 0.33 $\pm$ 0.08 \\
NGC5631       & -0.00529   & 0.128         & $<$ 7.90                 & 4.07       & 4.41        & 2.33        & 1150 $\pm$ 218  & 1080 $\pm$ 193  & 1160 $\pm$ 213  & 615 $\pm$ 154 \\  % 8.54 $\pm$ 0.43
NGC5720       & -0.00195   & 0.00651         & 9.22 $\pm$ 0.11                 & 4.29       & 4.90        & 2.28        & 2.86 $\pm$ 0.79  & 2.83 $\pm$ 0.78  & 3.23 $\pm$ 0.89  & 1.50 $\pm$ 0.40 \\
NGC5888       & -0.00332   & 0.0981         & 8.98 $\pm$ 0.12                 & 4.04       & 3.78        & 3.87        & 1.92 $\pm$ 0.59  & 1.78 $\pm$ 0.55  & 1.67 $\pm$ 0.52  & 1.71 $\pm$ 0.56 \\
NGC5929       & 0.000127   & -0.058         & 8.83 $\pm$ 0.29                 & 4.22       & 5.01        & 0.92        & 1.22 $\pm$ 0.13  & 1.18 $\pm$ 0.13  & 1.40 $\pm$ 0.15  & 0.26 $\pm$ 0.03 \\
NGC5954       & -0.0026   & 0.0669         & 8.44 $\pm$ 0.27                 & 3.83       & 3.79        & 2.70        & 0.85 $\pm$ 0.08  & 0.74 $\pm$ 0.08  & 0.73 $\pm$ 0.08  & 0.52 $\pm$ 0.06 \\
NGC6132       & -0.0071   & 0.0551         & 8.91 $\pm$ 0.22                 & 3.66       & 4.32        & 3.35        & 0.93 $\pm$ 0.22  & 0.78 $\pm$ 0.17  & 0.93 $\pm$ 0.20  & 0.72 $\pm$ 0.16 \\
NGC6150B      & -0.00555   & 0.0667         & 9.40 $\pm$ 0.26                 & 3.67       & 3.78        & 2.02        & 2.88 $\pm$ 0.35  & 2.42 $\pm$ 0.31  & 2.49 $\pm$ 0.33  & 1.33 $\pm$ 0.17 \\
NGC6154       & -0.00313   & 0.0795         & 8.81 $\pm$ 0.12                 & 3.95       & 4.07        & 3.71        & 2.50 $\pm$ 0.72  & 2.28 $\pm$ 0.67  & 2.35 $\pm$ 0.69  & 2.14 $\pm$ 0.57 \\
NGC6338       & -0.001   & -0.0277         & 9.16 $\pm$ 0.09                 & 3.97       & 4.76        & 3.29        & 18.10 $\pm$ 3.85  & 16.50 $\pm$ 3.46  & 19.70 $\pm$ 4.18  & 13.70 $\pm$ 2.90 \\
NGC6497       & -0.00368   & 0.0991         & 9.14 $\pm$ 0.18                 & 3.99       & 3.92        & 2.05        & 4.79 $\pm$ 0.99  & 4.40 $\pm$ 0.92  & 4.33 $\pm$ 0.90  & 2.27 $\pm$ 0.50 \\
UGC01659      & -0.00234   & 0.00211         & 9.09 $\pm$ 0.13                 & 4.37       & 4.84        & 2.45        & 1.74 $\pm$ 0.36  & 1.75 $\pm$ 0.37  & 1.94 $\pm$ 0.41  & 0.98 $\pm$ 0.25 \\
UGC01938      & -0.0021   & -0.032         & 8.90 $\pm$ 0.20                 & 4.33       & 5.26        & 3.59        & 1.21 $\pm$ 0.35  & 1.21 $\pm$ 0.35  & 1.46 $\pm$ 0.43  & 1.00 $\pm$ 0.28 \\
UGC02134      & -0.00259   & 0.0409         & 9.31 $\pm$ 0.05                 & 3.95       & 4.25        & 1.91        & 4.06 $\pm$ 0.51  & 3.69 $\pm$ 0.46  & 3.96 $\pm$ 0.49  & 1.78 $\pm$ 0.23 \\
UGC02222      & -0.0000146   & -0.0554         & 8.44  $\pm$ 0.21                 & 4.57       & 5.25        & 2.63        & 330.0 $\pm$ 60.4  & 346.0 $\pm$ 62.4  & 399.0 $\pm$ 72.3  & 199.0 $\pm$ 43.8 \\
UGC02239      & -0.00112   & -0.0195         & 8.80 $\pm$ 0.07                 & 4.07       & 4.76        & 4.97        & 3.14 $\pm$ 0.41  & 2.94 $\pm$ 0.40  & 3.43 $\pm$ 0.46  & 3.59 $\pm$ 0.57 \\
UGC03960      & -0.000335   & -0.0486         & $<$ 7.20                 & 4.20       & 4.89        & 9.25        & 34.60 $\pm$ 10.20  & 33.50 $\pm$ 9.34  & 39.00 $\pm$ 10.90  & 73.90 $\pm$ 21.40 \\  % 7.84 $\pm$ 0.39
UGC04136      & -0.000849   & -0.0188         & 8.48 $\pm$ 0.30                 & 4.04       & 4.64        & 10.70       & 12.90 $\pm$ 2.68  & 12.00 $\pm$ 2.49  & 13.70 $\pm$ 2.84  & 31.50 $\pm$ 6.19 \\
UGC04245      & -0.00392   & 0.0508         & 8.93 $\pm$ 0.17                 & 4.10       & 4.38        & 2.55        & 2.08 $\pm$ 0.38  & 1.96 $\pm$ 0.34  & 2.10 $\pm$ 0.37  & 1.22 $\pm$ 0.22 \\
UGC04262      & -0.002   & 0.0145         & 9.39 $\pm$ 0.24                 & 4.22       & 4.63        & 1.59        & 7.61 $\pm$ 1.56  & 7.38 $\pm$ 1.49  & 8.12 $\pm$ 1.66  & 2.77 $\pm$ 0.63 \\
UGC04659      & -0.00119   & -0.104         & 7.05 $\pm$ 0.19                 & 4.63       & 6.49        & 5.33        & 2.59 $\pm$ 0.52  & 2.77 $\pm$ 0.52  & 3.89 $\pm$ 0.73  & 3.19 $\pm$ 0.93 \\
UGC04730      & -0.000281   & -0.0814         & 7.55 $\pm$ 0.26                 & 4.05       & 5.38        & 5.53        & 0.04 $\pm$ 0.02  & 0.03 $\pm$ 0.01  & 0.05 $\pm$ 0.02  & 0.05 $\pm$ 0.02 \\
UGC05326      & 0.00105   & -0.212         & $<$ 6.68                 & 5.43       & 8.54        & 4.45        & 0.35 $\pm$ 0.09  & 0.44 $\pm$ 0.12  & 0.69 $\pm$ 0.18  & 0.36 $\pm$ 0.10 \\  
UGC05396      & -0.00102   & -0.035         & 8.18 $\pm$ 0.17                 & 4.56       & 5.26        & 3.58        & 0.96 $\pm$ 0.39  & 1.01 $\pm$ 0.41  & 1.16 $\pm$ 0.47  & 0.79 $\pm$ 0.26 \\
UGC08004      & -0.00271   & -0.0222         & 8.59 $\pm$ 0.16                 & 4.70       & 5.86        & 2.70        & 2.21 $\pm$ 0.69  & 2.40 $\pm$ 0.74  & 2.98 $\pm$ 0.92  & 1.38 $\pm$ 0.39 \\
UGC08231      & 0.00226   & -0.262         & 7.90 $\pm$ 0.23                 & 5.48       & 8.86        & 3.50        & 1.32 $\pm$ 0.24  & 1.67 $\pm$ 0.30  & 2.70 $\pm$ 0.48  & 1.07 $\pm$ 0.24 \\
UGC08234      & -0.00092   & -0.0075         & 8.73 $\pm$ 0.11                 & 3.97       & 4.32        & 6.99        & 324.0 $\pm$ 83.1  & 295.0 $\pm$ 74.3  & 321.0 $\pm$ 81.3  & 520.0 $\pm$ 135.0 \\
UGC08322      & -0.00178   & 0.0165         & 8.89 $\pm$ 0.17                 & 4.11       & 4.61        & 3.31        & 9.90 $\pm$ 2.62  & 9.36 $\pm$ 2.47  & 10.50 $\pm$ 2.74  & 7.52 $\pm$ 1.75 \\
UGC08733      & -0.00101   & -0.0859         & $<$ 8.05                 & 4.61       & 6.12        & 2.30        & 3.08 $\pm$ 0.51  & 3.27 $\pm$ 0.54  & 4.33 $\pm$ 0.71  & 1.63 $\pm$ 0.32 \\  
UGC08781      & -0.000444   & -0.0299         & 9.32 $\pm$ 0.10                 & 4.27       & 4.79        & 2.24        & 6.48 $\pm$ 1.56  & 6.36 $\pm$ 1.51  & 7.14 $\pm$ 1.69  & 3.34 $\pm$ 0.82 \\
UGC09598      & -0.00398   & 0.0602         & 8.80 $\pm$ 0.16                 & 4.24       & 4.59        & 3.00        & 3.01 $\pm$ 0.87  & 2.93 $\pm$ 0.86  & 3.17 $\pm$ 0.92  & 2.08 $\pm$ 0.53 \\
UGC09629      & -0.000801   & -0.0171         & 8.47  $\pm$ 0.13                 & 4.19       & 4.70        & 6.08        & 11.50 $\pm$ 2.54  & 11.00 $\pm$ 2.43  & 12.40 $\pm$ 2.75  & 16.00 $\pm$ 4.02 \\
UGC10097      & 0.00022   & -0.0642         & $<$ 8.63                 & 2.97       & 3.75        & 3.49        & 6.90 $\pm$ 1.77  & 4.71 $\pm$ 1.25  & 5.94 $\pm$ 1.59  & 5.52 $\pm$ 1.54 \\  
UGC10905      & -0.000319   & -0.0306         & 8.58  $\pm$ 0.12                 & 3.95       & 4.56        & 3.53        & 32.60 $\pm$ 8.58  & 29.60 $\pm$ 7.53  & 34.30 $\pm$ 8.83  & 26.50 $\pm$ 7.12 \\
CGCG163-062   & -0.000584   & -0.0603         & 7.90 $\pm$ 0.20                 & 4.63       & 5.49        & 3.55        & 1.33 $\pm$ 0.44  & 1.41 $\pm$ 0.47  & 1.67 $\pm$ 0.56  & 1.08 $\pm$ 0.32 \\
CGCG536-030   & -0.0014   & -0.0512         & 9.75 $\pm$ 0.26                 & 4.65       & 5.63        & 0.71        & 11.20 $\pm$ 1.10  & 11.90 $\pm$ 1.18  & 14.50 $\pm$ 1.43  & 1.83 $\pm$ 0.21 \\
IC0674        & -0.00283   & 0.0177         & 9.17  $\pm$ 0.16                 & 4.22       & 4.73        & 5.12        & 34.00 $\pm$ 7.54  & 33.00 $\pm$ 7.34  & 37.00 $\pm$ 8.27  & 40.10 $\pm$ 9.28 \\
IC3598        & -0.000973   & -0.0178         & 8.14 $\pm$ 0.16                 & 3.65       & 3.99        & 5.45        & 3.71 $\pm$ 1.53  & 3.11 $\pm$ 1.34  & 3.41 $\pm$ 1.47  & 4.65 $\pm$ 1.54 \\
Mrk1418       & 0.00509   & -0.276         & 6.18 $\pm$ 0.19                 & 5.37       & 8.43        & 2.21        & 0.19 $\pm$ 0.11  & 0.23 $\pm$ 0.13  & 0.37 $\pm$ 0.20  & 0.10 $\pm$ 0.04 \\  

\end{longtable*}

\begin{table*}
\centering
\caption{Statistics for \aco, $\log(M_\mathrm{mol})$, and $t_\mathrm{dep}$ \label{tab:statistics}}
\begin{tabular}{lcccccc}
  \hline\hline
  \multirow{2}{*}{Statistics} & \multicolumn{3}{c}{Mean $\pm$ Std} & \multicolumn{3}{c}{16$^\mathrm{th}$/50$^\mathrm{th}$/84$^\mathrm{th}$ Percentile Values} \\
  \cmidrule(lr){2-4} \cmidrule(lr){5-7}
  & MS  & GV  & RG  & MS  & GV  & RG \\
  \hline
  \aco(B13) [$\frac{\mathrm{M}_\odot}{\mathrm{K~km~s}^{-1}~\mathrm{pc}^2}$] & $4.29 \pm 0.52$ & $3.83 \pm 0.44$ & $4.19 \pm 0.23$ & 3.81/4.22/4.70 & 3.38/3.97/4.15 & 3.97/4.13/4.51 \\
  \aco(SL24) [$\frac{\mathrm{M}_\odot}{\mathrm{K~km~s}^{-1}~\mathrm{pc}^2}$] & $5.19 \pm 1.44$ & $4.33 \pm 0.48$ & $4.72 \pm 0.38$ & 3.88/4.79/6.19 & 3.89/4.56/4.74 & 4.33/4.63/5.21 \\
  \aco(T24*) [$\frac{\mathrm{M}_\odot}{\mathrm{K~km~s}^{-1}~\mathrm{pc}^2}$] & $2.93 \pm 1.33$ & $4.35 \pm 2.18$ & $4.68 \pm 2.46$ & 1.81/2.45/4.38 & 3.08/3.53/5.25 & 2.37/4.15/6.88 \\
  \midrule
  $\log(M_\mathrm{mol,MW}\ [\mathrm{M}_\odot])$ & $9.13 \pm 0.88$ & $9.27 \pm 0.49$ & $8.94 \pm 0.60$ & 8.27/9.45/9.88 & 8.98/9.45/9.64 & 8.41/9.09/9.47 \\
  $\log(M_\mathrm{mol,B13}\ [\mathrm{M}_\odot])$ & $9.13 \pm 0.84$ & $9.21 \pm 0.53$ & $8.93 \pm 0.58$ & 8.31/9.43/9.84 & 8.93/9.40/9.61 & 8.43/9.10/9.43 \\
  $\log(M_\mathrm{mol,SL24}\ [\mathrm{M}_\odot])$ & $9.20 \pm 0.80$ & $9.26 \pm 0.53$ & $8.98 \pm 0.58$ & 8.47/9.47/9.88 & 8.98/9.42/9.67 & 8.49/9.15/9.49 \\
  $\log(M_\mathrm{mol,T24*}\ [\mathrm{M}_\odot])$ & $8.92 \pm 0.77$ & $9.23 \pm 0.48$ & $8.91 \pm 0.53$ & 8.26/9.28/9.59 & 9.02/9.26/9.58 & 8.29/8.99/9.53 \\
  \midrule
  $t_\mathrm{dep}$(MW) [Gyr] & $2.69 \pm 2.43$ & $11.99 \pm 11.05$ & $264.1 \pm 356.4$ & 0.79/2.10/4.45 & 3.23/6.90/23.90 & 14.27/127.65/329.28 \\
  $t_\mathrm{dep}$(B13) [Gyr] & $2.59 \pm 2.36$ & $10.99 \pm 10.53$ & $253.5 \pm 335.1$ & 0.72/1.96/3.92 & 2.78/4.71/21.74 & 13.70/128.20/339.88\\
  $t_\mathrm{dep}$(SL24) [Gyr] & $2.99 \pm 2.82$ & $12.57 \pm 12.05$ & $279.0 \pm 360.2$  & 0.81/2.22/4.33 & 2.99/5.94/25.54 & 15.59/146.45/389.64 \\
  $t_\mathrm{dep}$(T24*) [Gyr] & $1.42 \pm 0.97$ & $12.64 \pm 13.02$ & $211.3 \pm 218.4$ & 0.35/1.22/2.41 & 2.40/5.52/28.50 & 21.42/136.45/481.84  \\
  \bottomrule
\end{tabular} 
%\tablecomments{}
\end{table*}

\begin{figure*}
\centering
\begin{minipage}{.4\linewidth}
\centering
\includegraphics[width=\linewidth]{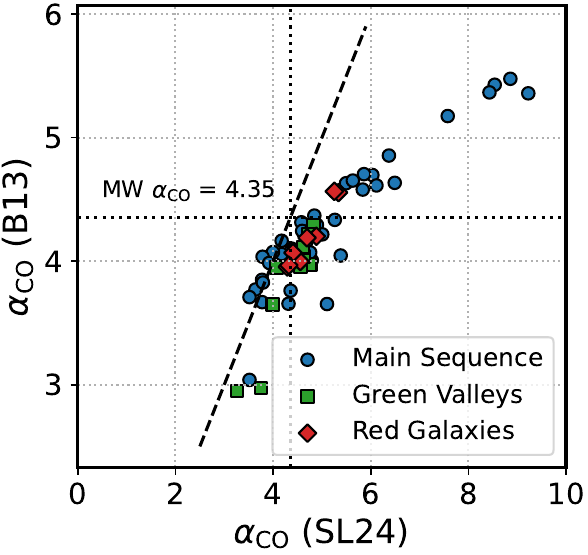}\\
%(a)
\end{minipage}
\hspace{0.2in}
\begin{minipage}{.37\linewidth}
\centering
\includegraphics[width=\linewidth]{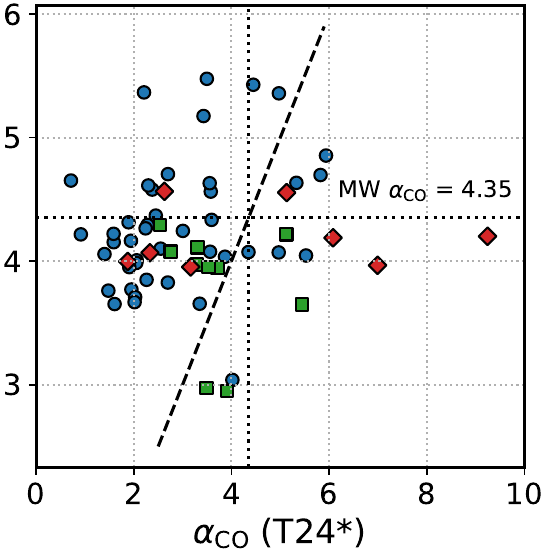}\\
%(b)
\end{minipage}
\caption{Comparison of \aco\ values estimated via different prescriptions (B13, SL24, and T24*, which represent Equations~\ref{eqn_alpha_B13}--\ref{eqn_aco_T24}). The dashed lines indicate a one-to-one relation, and the thick dotted lines label the Galactic \aco\ value of 4.35~$\mathrm{M_\odot\ (K~km~s^{-1}~pc^2)^{-1}}$.
While the predicted \aco\ values  do not agree well and show no systematic trend among different populations, the overall \aco\ variations are limited to a factor of two within and among prescriptions. The fractional \aco~uncertainty in these prescriptions is within $\pm$0.2~dex for T24* and $\pm$0.3~dex for B13 and SL24.}
\label{fig:sfr-aco-compare}
\end{figure*}

\subsubsection{Variable CO-to-H$_2$ Conversion Factors  \label{subsubsec:aco}} 

In addition to a constant MW \aco\ assumption as used in most previous studies, we also adopt three other \aco\ prescriptions in recent literature. The first two prescriptions are both functions of the normalized metallicity to the solar value ($Z'$) as well as the stellar mass surface density ($\Sigma_\mathrm{star}$). One of these prescriptions originates from \citet[hereafter B13]{co-to-h2}:
\begin{equation}
\alpha_\mathrm{CO} = 2.9\ \exp\left(\frac{0.4}{Z'}\right) \left(\frac{\Sigma_\mathrm{star}}{100\ \mathrm{M_\odot\ pc^{-2}}}\right)^{-\gamma}~,
\label{eqn_alpha_B13}
\end{equation}
where $\gamma = 0.5$ if $\Sigma_\mathrm{star} > 100\ \mathrm{M_\odot}$~pc$^{-2}$ and $\gamma = 0$ otherwise. Here we simplify the $\Sigma_\mathrm{total} \equiv \Sigma_\mathrm{star} + \Sigma_\mathrm{gas}$ term in B13 into $\Sigma_\mathrm{star}$ alone, as $\Sigma_\mathrm{star}$ is generally found to be an order of magnitude higher than $\Sigma_\mathrm{gas}$ in the EDGE-CALIFA sample \citep{2025A&A...699A.366C}. 
The other similar prescription is suggested by \citet[hereafter SL24]{2024ARA&A..62..369S}:
\begin{equation}
\alpha_\mathrm{CO} = 4.35\ (Z')^{-1.5}\ \left(\frac{\Sigma_\mathrm{star}}{100\ \mathrm{M_\odot\ pc^{-2}}}\right)^{-\gamma}~,
\label{eqn_alpha_SL24}
\end{equation}
where $\gamma = 0.25$ if $\Sigma_\mathrm{star} > 100\ \mathrm{M_\odot}$~pc$^{-2}$ and $\gamma = 0$ otherwise \citep{2024ApJ...964...18C}. 
Both the B13 and SL24 prescriptions suggest at least a $\pm$0.3~dex uncertainty, based on the \aco~measurements adopted in their calibrations.

To distinguish from Equations~\ref{eqn_alpha_B13} and~\ref{eqn_alpha_SL24} which rely on indirect (or non-CO) tracers, we also implement a new \aco\ prescription based on inclination-corrected CO velocity dispersion \citep[][and in preparation, hereafter T24*]{2024ApJ...961...42T}:
\begin{equation}
\log \alpha_\mathrm{CO} = -0.96\ \log (\Delta v)_\mathrm{2kpc} + 1.77~, 
%old: y = -0.56x + 1.19
\label{eqn_aco_T24}
\end{equation}
where $(\Delta v)_\mathrm{2kpc}$ is measured at 2-kpc scales, in units of km~s$^{-1}$, and corrected for galaxy inclination by applying a $\sqrt{\cos i}$ factor \citep[see Appendix in][]{2022AJ....164...43S}. This prescription (Equation~\ref{eqn_aco_T24}) is a modified version with slightly different coefficients from Equation~2 in \citet{2024ApJ...961...42T}, where they use $(\Delta v)_\mathrm{150pc}$ instead of $(\Delta v)_\mathrm{2kpc}$. 
In a follow-up work (Y.-H. Teng et al. 2026, in preparation), it is found that the anti-correlation between \aco\ and $\Delta v$ can be extended up to 2-kpc scales with a $< 0.2$~dex scatter, because the value of $(\Delta v)_\mathrm{2kpc}$ is still dominated by $(\Delta v)_\mathrm{150pc}$, thereby leading to strong correlations among cross-scale $\Delta v$ across the same galaxy sample as in \citet{2024ApJ...961...42T}.
Therefore, here we apply the modified $(\Delta v)_\mathrm{2kpc}$-based prescription to match our GBT resolutions of $\sim$2 kpc.

To obtain $Z'$ required by Equations~\ref{eqn_alpha_B13} and~\ref{eqn_alpha_SL24}, we assume $12+\log(\mathrm{O/H})_\odot = 8.69$ \citep{2009ARA&A..47..481A} and employ the O3N2 metallicity calibration from \citet{2017MNRAS.465.1384C}.
We note that the choice of metallicity calibration can vary the derived metallicity values by a factor of 2--3 \citep{2019A&A...623A...5D,2024ApJ...961...42T}, and both the B13 and SL24 \aco~prescriptions are sensitive to the choice of calibration. 
In B13, the metallicity-dependent term in exponential form is motivated by theoretical derivations \citep{2010ApJ...716.1191W}, while the power-law metallicity dependence in SL24 is based on observational measurements using various metallicity calibrators, including \citet{2004MNRAS.348L..59P} and the \citet{2016MNRAS.457.3678P} S-calibration.
However, we find that using those calibrations result in sub-solar metallicity values for the majority of our sample, which would predict unrealistically high \aco~values based on B13 and SL24. On the other hand, the \citet{2017MNRAS.465.1384C} calibration is 
widely tested and accounts for corrections from previous calibration methods \citep[see review by][]{2019A&ARv..27....3M}, and it results in near-solar metallicity for most galaxies in our sample.   

Therefore, based on \citet{2017MNRAS.465.1384C}, we derive $Z'$ on a pixel-by-pixel basis, using [OIII], [NII], H$\alpha$, and H$\beta$ line maps in the \texttt{Pipe3D} data products. We compute $\log(\mathrm{O/H})$ values only for pixels that fulfill the criteria of being compatible with star formation ionization (i.e., H$\alpha$ equivalent width $> 6\mathrm{\AA}$ and S/N $>1$ for all emission lines involved in the classical O3-N2-BPT diagram, below the \citealt{2001ApJ...556..121K} curve).
As the resulting $Z'$ measurements can be sparse in many regions of galaxies, we fit a $Z'$ radial gradient for each galaxy to infer a $Z'$ value for every pixel:
\begin{equation}
\log Z' = a \cdot \log R_\mathrm{gal} + b~,
\label{eqn_Zprime_fit}
\end{equation}
where $Z'$ is in units of $Z_\odot$ and $R_\mathrm{gal}$ is the de-projected galactocentric radius in arcseconds. The best-fit $a$ and $b$ values for each galaxy are listed in Table~\ref{tab:derived}. 
The resolved $Z'$ map together with the $\Sigma_\mathrm{star}$ map from CALIFA then allows us to obtain a resolved \aco\ map for each galaxy. 
As for implementing Equation~\ref{eqn_aco_T24}, we simply use the derived CO velocity dispersion maps and multiply them with a $\sqrt{\cos i}$ factor to correct for galaxy inclination \citep{2022AJ....164...43S,2024ApJ...961...42T}.      

To obtain a spatially-weighted global \aco\ value for each galaxy, our procedure is as follows. First, we compute \aco\ values pixel by pixel based on the GBT maps using Equations~\ref{eqn_alpha_B13}--\ref{eqn_aco_T24}, respectively. Next, we multiply these derived \aco\ maps by the moment-0 maps (see Figure~\ref{fig:mom0-datapref} and Appendix~\ref{appendix:maps}) to obtain resolved $M_\mathrm{mol}$ maps. Then, we compute the integrated $M_\mathrm{mol}$ and CO intensity by summing both the moment-0 maps and the derived $M_\mathrm{mol}$ maps over regions within the R$_{25}$ radius. Lastly, we obtain a global \aco\ value for each galaxy by dividing the total $M_\mathrm{mol}$ by the integrated CO intensity. The derived \aco\ values and their statistics under different prescriptions are listed in Tables~\ref{tab:derived} and~\ref{tab:statistics} .

Figure~\ref{fig:sfr-aco-compare} presents comparisons between our derived \aco\ values for all galaxies using Equations~\ref{eqn_alpha_B13} (B13), \ref{eqn_alpha_SL24} (SL24), and \ref{eqn_aco_T24} (T24*), respectively. 
Comparing between B13 and SL24 that are both metallicity dependent, we find consistent predictions only for MS galaxies with near-solar metallicities, which show \aco\ values slightly lower than the Galactic value of 4.35 (i.e., the blue points that align well with the 1-to-1 relation in the left panel of Figure~\ref{fig:sfr-aco-compare}). 
For galaxies with sub-solar metallicities, the power-law term in SL24 can easily lead to higher \aco\ values than the exponential term in B13, which explains the deviation from the 1-to-1 relation beyond the Galactic \aco. On the other hand, the $\Sigma_\mathrm{star}$ term in B13 has a steeper slope than that in SL24, explaining why \aco(B13) is lower for many non-MS galaxies that typically have higher stellar masses and densities. 
Finally, as shown in the right panel of Figure~\ref{fig:sfr-aco-compare}, T24* depends purely on the CO velocity dispersion and predicts $\sim$2 times lower-than-Galactic \aco\ for most MS galaxies. For GV and RG, T24* gives similar \aco\ predictions to B13 and SL24, all showing a mean \aco\ $\sim$4 (see Table~\ref{tab:statistics}).
We note that UGC~04136 is excluded in the figure, as its edge-on inclination ($i\sim90^\circ$) would lead to unrealistically high \aco\ estimates due to the $\sqrt{\cos i}$ correction of $(\Delta v)_\mathrm{2kpc}$ in Equation~\ref{eqn_aco_T24}.   

Overall, the tested \aco\ prescriptions predict diverse \aco\ distributions across our galaxy sample, and no systematic \aco\ dependence is found for different galaxy types. These \aco\ variations are typically within a factor of 2--3 among different prescriptions (see also Table~\ref{tab:statistics}). For MS galaxies, the two $Z'$-dependent prescriptions (B13 and SL24) tend to give higher \aco\ than the CO-based prescription (T24*). This is partially because of significant $Z'$ variations across galaxies which do not affect T24*, but it also suggests that CO velocity dispersion as a ``starburst emissivity term'' \citep[e.g.,][]{2024ARA&A..62..369S,2025ApJ...994..263S} in \aco\ predictions tends to give lower values than using $\Sigma_\mathrm{star}$ in B13 and SL24. 
Compared to MS galaxies, \aco\ predictions for GV and RG are more consistent among the three prescriptions.  
We caveat that these \aco\ prescriptions were developed based on MS galaxies, and thus they might not be appropriate for quenched systems below the MS. 
However, applying these prescriptions is the best approach currently available.
To obtain reliable \aco\ predictions for GV and/or RGs, systematic and spatially resolved \aco\ measurements in such environments are needed.

\begin{figure*}
\begin{minipage}{.43\linewidth}
\centering
\includegraphics[width=\linewidth]{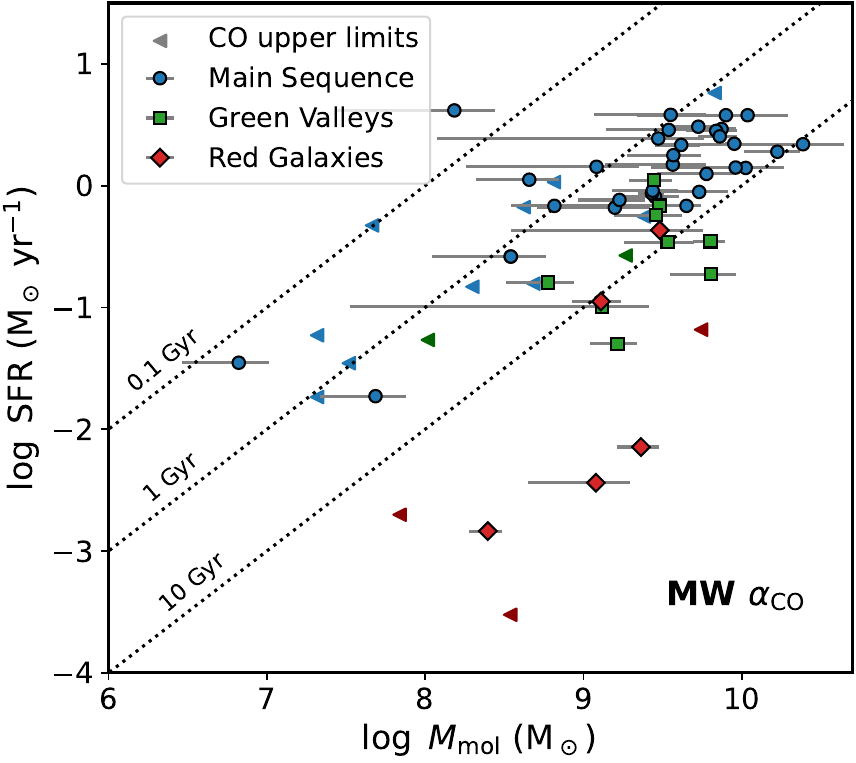}\\
(a)
\end{minipage}
\hfill
\begin{minipage}{.555\linewidth}
\centering
\includegraphics[width=\linewidth]{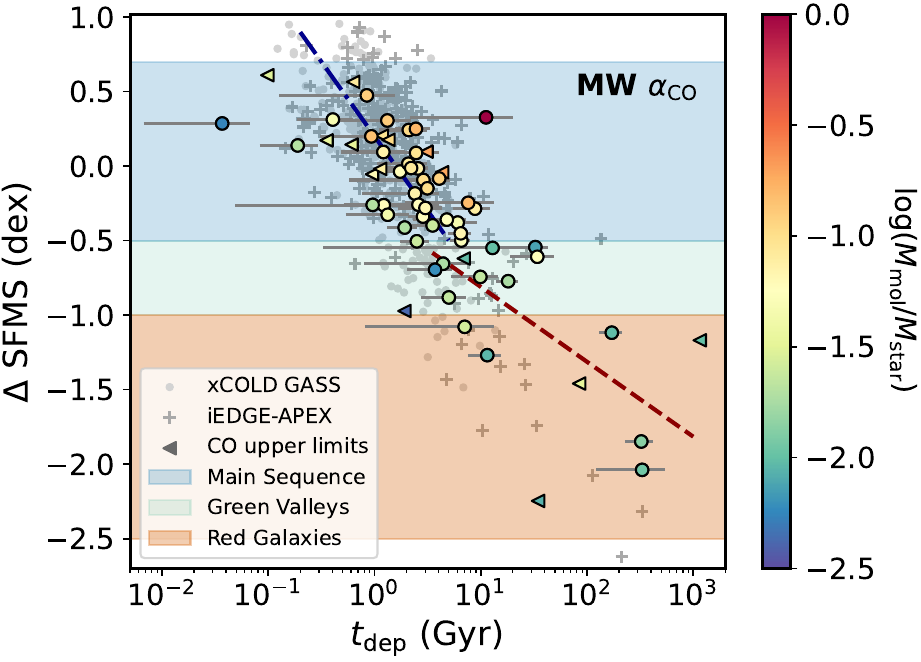}\\
(b)
\end{minipage}
\caption{(a) The SFR--$M_\mathrm{mol}$ relation across all \Ngal galaxies, using H$\alpha$-based SFR estimates and $M_\mathrm{mol}$ derived via a Galactic \aco. The dotted lines show constant molecular gas depletion times ($t_\mathrm{dep}$) of 0.1, 1, and 10 Gyr. (b) The derived $t_\mathrm{dep}$ increases systematically as galaxies go from MS to GV and to RG, suggesting that low SFRs in quenched galaxies are mostly caused by suppressed star formation efficiency (SFE) rather than the lack of molecular gas. 
The light gray points and pluses are from the xCOLD GASS survey \citep{2017ApJS..233...22S} and the iEDGE APEX data \citep{2025A&A...699A.366C}, respectively, which align with an expected slope of -1 (blue, dashed-dotted line).
Our GV and RG sample reveals an increased $t_\mathrm{dep}$ with a best-fit slope of -0.5 (red, dashed line).
The color-coded molecular-to-stellar mass ratio shows higher values in MS but remains similarly low across GV and RGs, implying a dominant SFE-driven quenching from the GV to RG populations. 
}
\label{fig:scalings}
\end{figure*}

\subsection{Gas Depletion and Star Formation Efficiency  \label{subsec:sfe}} 

Based on the derived star formation rates (SFR) and molecular gas masses ($M_\mathrm{mol}$) in Section~\ref{subsec:estimation}, we investigate the SFR--$M_\mathrm{mol}$ relation \citep{1996AJ....112.1903Y} and study the molecular gas depletion time ($t_\mathrm{dep}$; or star formation efficiency SFE = $t_\mathrm{dep}^{-1}$) across our galaxy sample. 

Figure~\ref{fig:scalings}a shows the SFR--$M_\mathrm{mol}$ relation across all \Ngal galaxies, assuming a constant MW \aco. The MS galaxies span a wide range of $M_\mathrm{mol}$ from $10^7-10^{10}$ M$_\odot$, and their typical $t_\mathrm{dep}$ value is around 2~Gyr. On the other hand, all GV and RGs are found to have $M_\mathrm{mol} \gtrsim 10^8$ M$_\odot$, which suggests even larger molecular gas reservoirs than some MS galaxies. Thus, combined with the low SFRs for GVs and RGs, such a good amount of molecular gas in those galaxies leads to significantly longer $t_\mathrm{dep}$ than most of the MS galaxies. It is also clear from Figure~\ref{fig:scalings}a that most RGs have substantially longer $t_\mathrm{dep}$ beyond a few tens of Gyr, causing a clear gap from all other galaxies. 

To verify whether $t_\mathrm{dep}$ changes systematically across MS, GV, and RG populations, we examine the relation of $t_\mathrm{dep}$ with the galaxy's offset from SFMS, defining that $\Delta$SFMS $= \log$(SFR/SFR$_\mathrm{MS}$), where SFR$_\mathrm{MS}$ is specified in Equation~\ref{eqn_sfms}. As shown in Figure~\ref{fig:scalings}b, we find a steady increase of $t_\mathrm{dep}$ as $\Delta$SFMS decreases, which clearly deviates from the scenario of a constant gas depletion and indicates more amount of molecular gas than expected in quenched galaxies. 
We note that while $\Delta$SFMS and $t_\mathrm{dep}$ are both functions of SFR, their intrinsic anti-correlation can only cause a linear change between $\Delta$SFMS and $\log(t_\mathrm{dep})$, which is not sufficient to explain the 4~dex increase of $t_\mathrm{dep}$ over just 2.5~dex of SFR range in Figure~\ref{fig:scalings}b, and thus SFE must play a significant role in this systematic change with $\Delta$SFMS.

By comparison, the detections in the xCOLD GASS survey \citep[i.e., gray points in the background of Figure~\ref{fig:scalings}b;][]{2017ApJS..233...22S}, shows a roughly linear trend that could be caused by the intrinsic correlation between $\Delta$SFMS and $t_\mathrm{dep}$ (see the blue dashed-dotted line in Figure~\ref{fig:scalings}b which has a slope of -1). These data points are based on aperture-corrected measurements reported in \citet[Table 3]{2017ApJS..233...22S}, using a MW \aco\ to ensure a consistent comparison with our data.
Our MS sample aligns well with that of xCOLD GASS, which suggests a consistent SFE among MS galaxies. While xCOLD GASS also includes a small number of detected GV and RGs, the iEDGE survey using the Atacama Pathfinder Experiment telescope (APEX) detected a larger sample of GV and RGs at S/N $> 5$ (gray pluses in Figure~\ref{fig:scalings}b; \citealt{2020A&A...644A..97C,2025A&A...699A.366C}), and they found a generally longer $t_\mathrm{dep}$ in those galaxies.
To compare with our galaxy-integrated measurements, here we extract the aperture corrected global quantities (SFR, $M_\mathrm{star}$, $M_\mathrm{mol}$, and SNR) provided in \citet{2025A&A...699A.366C} and apply a cut at S/N = 5. 

Overall, we find our GV and RG sample agrees well the iEDGE APEX survey, and three of our RGs (NGC~5216, UGC~02222, and UGC~08234) show an even further increase in $t_\mathrm{dep}$ beyond 100~Gyr.
We have checked that such long $t_\mathrm{dep}$ in these three galaxies is unlikely to be caused by underestimated uncertainties, as we have done an additional `fake source' test (as described in Section~\ref{subsec:maps}) customized to the detected CO features in those galaxies and obtained similar uncertainties within 40\%. 
The best-fit relation for our detected GV/RG sample is $\Delta\mathrm{SFMS} = -0.5 \log(t_\mathrm{dep}) - 0.3$ (i.e., red dashed line in Figure~\ref{fig:scalings}b), indicating a substantial drop in SFE compared to the MS sample with a slope of -1.
In summary, our finding of a significant $t_\mathrm{dep}$ increase across GV and RGs suggests that the star formation quenching process in the galaxies detected in CO is primarily driven by a decline in SFE rather than molecular gas exhaustion (see Section~\ref{sec:discussion} for further discussions). 

In Figure~\ref{fig:scalings}b, the color coding shows that there is a clear drop in the  molecular-to-stellar mass ratio ($M_\mathrm{mol}/M_\mathrm{star}$) from MS galaxies (where the median $\log[M_\mathrm{mol}/M_\mathrm{star}]\sim-1$) to those below MS (where $\log[M_\mathrm{mol}/M_\mathrm{star}]\sim-2$). This suggests that in addition to SFE effects, the reduction in molecular gas plays an important role for the transition from MS to GV. On the other hand, the transition from GV to RG  shows no accompanying drop in $M_\mathrm{mol}/M_\mathrm{star}$, which suggests that this evolution is dominated by a decrease in SFE.
Furthermore, for MS galaxies we observe a systematic change in $M_\mathrm{mol}/M_\mathrm{star}$ across (orthogonal to) the $\Delta$SFMS$-t_\mathrm{dep}$ relation, with higher $M_\mathrm{mol}/M_\mathrm{star}$ values (red-orange) on the right side and lower (yellow-green) on the left side of the main relation. This change is due to the intrinsic correlation of the $M_\mathrm{mol}/M_\mathrm{star}$ ratio with $\Delta$SFMS ($\sim$SFR/$M_\mathrm{star}$) and $t_\mathrm{dep}$, indicating that the SFE (SFR/$M_\mathrm{mol}$) shows little variation across the population of MS galaxies --- because large variations would cancel the correlation \citep{2016MNRAS.462.1749S,2017ApJS..233...22S,2020A&A...644A..97C,2025A&A...699A.367C,2025A&A...699A.366C}.
However, this same correlation disappears for the detected galaxies below the MS, which means that SFE must vary substantially across our detected GV and RG sample. 
In Appendix~\ref{appdendix:sfe}, we show that \aco\ choices have no impact on the qualitative results in Figure~\ref{fig:scalings}.

\begin{figure*}
\centering
\begin{minipage}{.5\linewidth}
\centering
\includegraphics[width=\linewidth]{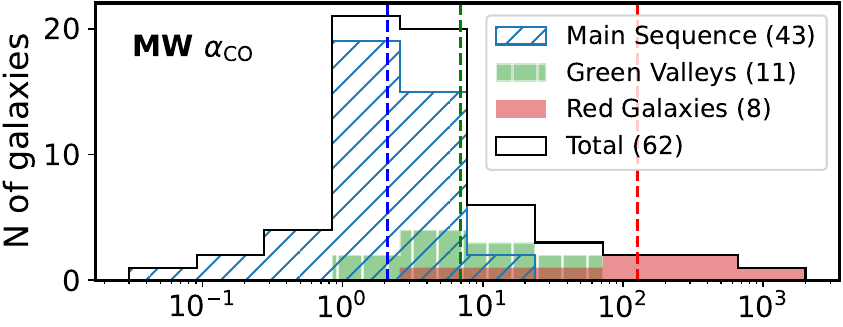}\\
\end{minipage}
\hfill
\begin{minipage}{.48\linewidth}
\centering
\includegraphics[width=\linewidth]{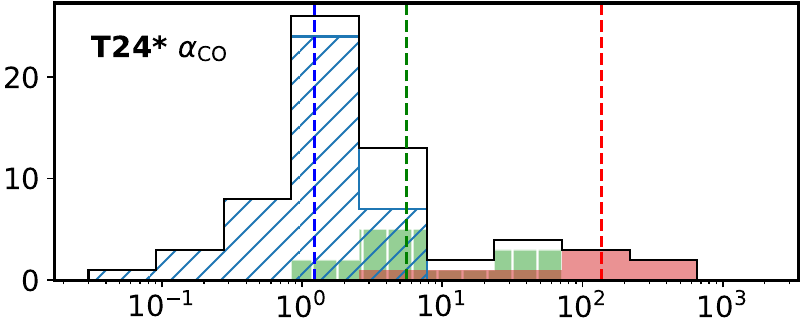}\\
\end{minipage}
\\
\begin{minipage}{.5\linewidth}
\centering
\includegraphics[width=\linewidth]{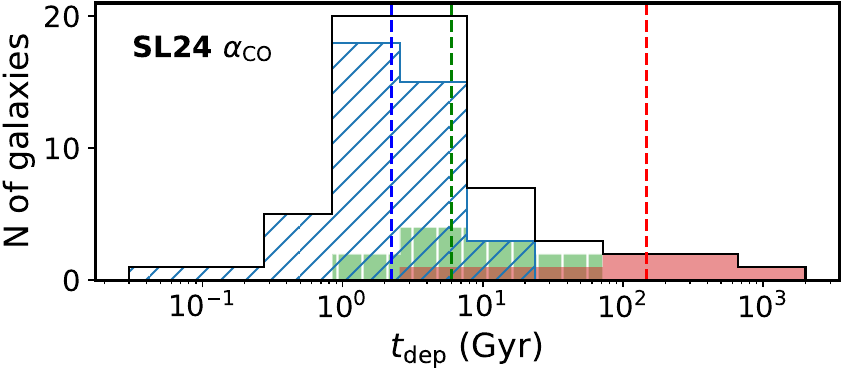}\\
\end{minipage}
\hfill
\begin{minipage}{.48\linewidth}
\vspace{2.5ex}
\hspace{0.3ex}
\includegraphics[width=\linewidth]{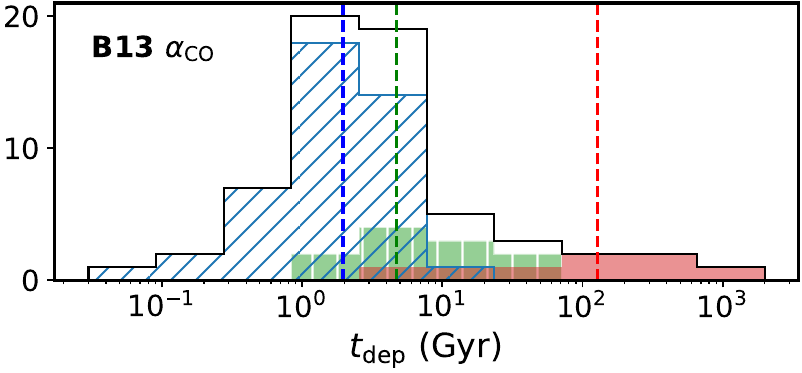}\\
\end{minipage}
\caption{Histograms of the derived gas depletion times under four different \aco\ treatments: constant MW value, a CO velocity dispersion-based prescription (T24*; Equation~\ref{eqn_aco_T24}), and two metallicity + stellar mass density-based prescriptions (B13 and SL24; Equation~\ref{eqn_alpha_B13} and \ref{eqn_alpha_SL24}). The vertical dashed lines represent the median $t_\mathrm{dep}$ values for each galaxy population. All four prescriptions reveal distinct $t_\mathrm{dep}$ distributions for MS, GV, and RG groups, showing longer $t_\mathrm{dep}$ in more quenched galaxies.}
\label{fig:hist-tdep}
\end{figure*}

Figure~\ref{fig:hist-tdep} compares the derived $t_\mathrm{dep}$ distributions and values using different \aco\ prescriptions. 
We find that all four prescriptions reveal a gradual increase in the $t_\mathrm{dep}$ range from the MS to GV and an even more substantial increase toward RGs. 
In Table~\ref{tab:statistics}, we report the mean, standard deviation, median, and the 16th and 84th percentile values of $t_\mathrm{dep}$ for each galaxy type.
With the MW \aco, the median $t_\mathrm{dep}$ derived for MS, GV, and RGs are $2.10^{+2.35}_{-1.31}$, $6.90^{+17.00}_{-3.67}$, and $127.7^{+201.6}_{-113.4}$ Gyr, respectively. Other prescriptions also follow a similar trend.
For all galaxy types, the differences in $t_\mathrm{dep}$ among these prescriptions are generally within a factor of 2, while there is a tendency for SL24 to predict longer $t_\mathrm{dep}$ and T24* to predict shorter $t_\mathrm{dep}$ for MS galaxies due to the nature of those prescriptions (see Section~\ref{subsubsec:aco}).

\section{Discussion  \label{sec:discussion}} 

Our new CO\,(1--0) nearby galaxy survey, GBT-EDGE, has revealed a comparable amount of molecular gas across a sample of galaxies at different evolutionary stages, with galaxy-integrated molecular gas masses spanning from $10^{8-10}$ M$_\odot$ for either main sequence (MS), green valley (GV), or red galaxies (RGs). The CO datasets, combined with existing optical IFU measurements, further uncover a smooth and significant increase in molecular gas depletion time ($t_\mathrm{dep}$) as galaxies evolve from MS toward quenched stages. These results suggest that the decline of star formation in quenched galaxies is mainly due to reduced star formation efficiency (SFE) of their molecular gas, rather than a lack of molecular gas reservoir in those galaxies. 

\subsection{SFR and \aco\ Uncertainties  \label{subsec:uncertainty}} 

The scenario of suppressed SFE in quenched galaxies strengthens when we account for possible biases or uncertainties in the estimation of SFR and molecular gas mass. Inferring SFRs in quenched galaxies, either using H$\alpha$ or stellar synthesis analyses only provides an upper limit of SFR, as H$\alpha$ emission can be contributed by ionization sources unassociated with star formation \citep{2007ApJS..173..267S,2012ARA&A..50..531K,2015A&A...584A..87C,2020MNRAS.492.3073L}.
With our SFR estimates being upper limits, all the derived $t_\mathrm{dep}$ are therefore lower limits, indicating that the long $t_\mathrm{dep}$ we found in quenched galaxies may be even longer, which remains supportive of the reduced SFE scenario.

The major uncertainty in molecular gas mass estimation arises from the variation of the CO-to-H$_2$ conversion factor, \aco. 
While there is a lack of systematic \aco\ measurements for quenched galaxies, recent studies using MS-calibrated \aco\ prescriptions have suggested lower \aco\ values for galaxies below the MS \citep{2017MNRAS.470.4750A,2024ARA&A..62..369S,2025ApJ...994..263S}.  
By implementing four \aco\ prescriptions from current literature (Equations~\ref{eqn_alpha_B13}--\ref{eqn_aco_T24} and MW \aco), we find that galaxy-integrated \aco\ variations are overall limited to a factor of 2 across our sample (Figure~\ref{fig:sfr-aco-compare} and Table~\ref{tab:derived}), which is incapable of explaining the $> 2$~dex change in $t_\mathrm{dep}$ spanned by our GV and RG sample (Figures~\ref{fig:scalings} and~\ref{fig:hist-tdep}). 
We note that our RG sample showing $M_\mathrm{mol} \gtrsim 10^8$~M$_\odot$ is also consistent with previous studies of CO-rich early-type galaxies at similar distances \citep[e.g.,][]{2011MNRAS.414..940Y,2013MNRAS.432.1796A}.  

In addition, we do not observe a significant difference in the global \aco\ values among MS, GV, and RG groups using any of the prescriptions (Table~\ref{tab:statistics} and Figure~\ref{fig:sfr-aco-compare}). This is contrary to the result in \citet{2025ApJ...994..263S}, where they apply the SL24 prescription to local massive galaxies and find systematically lower global \aco~values for galaxies below the MS. As discussed in \citet{2025ApJ...994..263S}, this discrepancy may be due to their weighting being biased toward the central regions in those galaxies.   
To further confirm if \aco~varies systematically with galaxy's offset from the SFMS, future molecular gas and \aco~measurements on a larger sample of quenched galaxies will be crucial.

\subsection{The Drivers of Reduced SFE  \label{subsec:sfe-drivers}}  

Our finding of a systematic $t_\mathrm{dep}$ (or SFE) variation with the offset from SFMS (i.e., $\Delta$SFMS, or ``specific SFRs'' as used in some studies) is consistent with previous galaxy-integrated molecular gas surveys in the low-redshift Universe \citep{2015ApJ...800...20G,2017ApJS..233...22S,2020A&A...644A..97C,2020ApJ...903..145L,2025ApJ...978...23B}.
In particular, \citet{2020A&A...644A..97C} also suggests that low SFE (instead of low molecular gas fraction) dominates the quenching from GV to RG, which is in line with our result in Figure~\ref{fig:scalings}b.
Moreover, previous studies on the SFR--$M_\mathrm{mol}$ relation of quenched galaxies, including early-type, GV, and RGs, have reported a significantly lower intercept than that of star-forming galaxies \citep{2014MNRAS.444.3427D,2022ApJ...926..175L,2025A&A...699A.367C}, showing lower SFE than MS galaxies as indicated in Figure~\ref{fig:scalings}a.  

Recent spatially-resolved studies have further shown similar trends to those in Figure~\ref{fig:scalings}, particularly in the inner regions of galaxies where SFE is found to be suppressed \citep{2023ApJ...952..122G,2024ApJ...964..120P,2024ApJ...962...88V,2025A&A...699A.366C}. While these studies tend to classify SFE-driven or gas-driven quenching within different regions of galaxies, it has been shown that the resolved and global classifications on quenching modes are generally in good agreement, and that a low global SFE value can be a strong indicator of SFE-dominated quenching \citep{2026ApJ...999..263L}.

Given various supporting evidence for reduced SFE in quenched galaxies, some questions remain: What prevents the molecular gas from forming stars? How do different physical processes control SFE and drive the quenching of active disk galaxies? 
While it is known that galaxy interactions such as ram pressure stripping or frequent high-speed encounters can cease star formation by removing gas from galaxies \citep[i.e., environmental quenching;][]{1996Natur.379..613M,1999MNRAS.308..947A,2010ApJ...721..193P,2022ApJ...940..176V,2023ApJ...952..122G}, our results suggest that reduced SFE can play a more dominant role than gas removal in driving the low SFR of quenched galaxies (Figure~\ref{fig:scalings}b).
Additionally, the only three galaxies in close interacting pairs (NGC~0169, 5929, and 5954) in our sample also span a wide range of $t_\mathrm{dep}$ (see Table~\ref{tab:derived}), while their $M_\mathrm{mol}/M_\mathrm{star} \sim 0.1$ are similar to the average of our MS sample.

Alternatively, feedback processes from AGNs may also disrupt surrounding gas via radiative or shock-induced heating, thereby preventing the gas from gravitational collapse \citep{2019MNRAS.487.1823L,2020MNRAS.492.3073L,2023ApJ...944..108B}. In our sample, however, only two galaxies are reliably identified to host a weak or strong AGN \citep[NGC~5216 and 5929;][]{2021A&A...648A..64K}. For NGC~5216 (classified as an RG), only a low amount of CO is located on its galaxy disk (Figure~\ref{fig:moments-datapref}), and thus AGN feedback is unlikely to have a significant impact on that gas. A recent study on the EDGE-CALIFA sample has also reported limited effects of AGN feedback on quenched galaxies \citep{2025A&A...697A.149B}. 

Morphological quenching is another possibility to prevent star formation, which can happen even with substantial amount of gas being present \citep{2009ApJ...707..250M}. Such process includes the development of bars, bulges, or spheroids to stabilize galaxies' gas disks \citep{2012ApJ...758...73S}, and thus morphological quenching is found to be critical in galaxy centers and provides evidence for an inside-out galaxy quenching scenario \citep{2019ApJ...872...50L,2021A&A...648A..64K,2022MNRAS.514.5035L,2023ApJ...943....7M,2024ApJ...964..120P,2025ApJ...978...23B}.   
It is also possible that CO\,(1--0) is tracing gas that is not dense enough to form stars, since its critical density is only $\sim 10^3$~cm$^{-3}$ \citep{2015PASP..127..299S,2022ApJ...925...72T}. 
If our CO detections mostly come from low-density gas, it could explain why the gas does not form stars efficiently, and why the derived SFE based on CO is low.

Given that we observe CO emission near $R_{25}$ in many galaxies (Figure~\ref{fig:moments-datapref}), morphological quenching likely has limited influence in those regions. As that gas lies in the outer disks, it could be relatively diffuse, compared to typical star-forming galaxies with molecular gas concentration toward the centers of galaxies. 
To further examine if our observed CO indeed traces low-density gas and leads to underestimated SFE in quenched galaxies, future observations with dense gas tracers can be helpful \citep[e.g., HCN;][]{2024ApJ...963..115L,2025A&A...693L..13N}. Resolved studies with constraints from multiple CO isotopologue lines \citep[e.g.,][]{2023ApJ...950..119T} in GV and RGs will also be critical to discern what specific gas conditions regulate SFE and how they affect galaxy quenching at various evolutionary stages.

\section{Conclusions  \label{sec:conclusion}} 

We present GBT-EDGE, a new CO\,(1--0) survey across \Ngal nearby massive galaxies, covering a sample of 43 main sequence (MS), 11 green valley (GV), and 8 red galaxies (RGs) in the local Universe with stellar masses above $3\times10^8$~M$_\odot$. By combining the CO observations with optical IFU data from the CALIFA survey, we estimate star formation efficiencies (SFE) across the sample and study possible mechanisms for galaxy quenching. Our main results are summarized as follows: 

\begin{enumerate}

\item We produce moment maps for all galaxies, using signal masks based on CO data dilation and constraints from H$\alpha$ velocity field. The moment-0 images reveal diverse and extended molecular gas structures across entire galaxies, including some galaxies with significant emission in their outer radii. The moment-1 images generally show clear rotation features for well-detected disk galaxies. 

\item We determine global SFRs via two approaches: 1) extinction-corrected H$\alpha$ line fluxes and 2) star formation history within an age of 33 Myr. We find that both methods give consistent results for MS and GV galaxies (Figure~\ref{fig:sfr-aco-compare}a). For RGs where both are likely overestimates, using star formation history leads to substantially overestimated SFRs (even higher than using H$\alpha$), likely due to large uncertainties in the luminosity fractions assigned to different age bins.

\item We derive and compare galaxy-integrated molecular gas masses ($M_\mathrm{mol}$) using different assumptions for the CO-to-H$_2$ conversion (Table~\ref{tab:derived}), including prescriptions from \citet{co-to-h2}, \citet{2024ARA&A..62..369S}, and \citet{2024ApJ...961...42T}. In addition, we employ metallicity calibration from \citet{2017MNRAS.465.1384C} and compute a best-fit metallicity gradient for each galaxy.
Overall, the global \aco\ values and thus $M_\mathrm{mol}$ are found to vary by a factor of 2 within and among prescriptions (Figure~\ref{fig:sfr-aco-compare} and Table~\ref{tab:statistics}). We obtain comparable $M_\mathrm{mol}$ spanning $10^{8-10}$~M$_\odot$ across our sample, regardless of galaxy types (MS, GV, or RG). 

\item The SFR--$M_\mathrm{mol}$ relation reveals a distinct gap between detected RGs and other galaxies, indicating a substantially longer gas depletion time ($t_\mathrm{dep}$) in such retired environments (Figure~\ref{fig:scalings}a). Assuming a Galactic \aco, the median $t_\mathrm{dep}$ for MS/GV/RG is $2.10^{+2.35}_{-1.31}$ / $6.90^{+17.00}_{-3.67}$ / $127.7^{+201.6}_{-113.4}$ Gyr, respectively. The gradual increase of $t_\mathrm{tep}$ from MS to GV and to RG is consistently shown across all \aco\ choices (Figure~\ref{fig:hist-tdep}).

\item We find that $t_\mathrm{dep}$ increases systematically with the distance from the MS ($\Delta$SFMS) across our entire sample (Figure~\ref{fig:scalings}b). Both the molecular-to-stellar mass ratio ($M_\mathrm{mol}$/$M_\mathrm{star}$) and SFE drop as galaxies transit from MS to GV. However, for GV and RGs, neither does $M_\mathrm{mol}$ nor $M_\mathrm{mol}$/$M_\mathrm{star}$ show correlation with $\Delta$SFMS. These results show that for CO-detected objects, galaxy quenching from GV to RG is primarily driven by low SFE rather than a deficit in molecular gas.

\item As CO is detected near the $R_{25}$ radius in many quenched galaxies, we do not expect morphological quenching or AGN feedback to be the dominant quenching mechanisms for our sample. We suspect that low gas density can be a major reason for reduced SFE, but additional observations of dense gas tracers such as HCN or CO isotopologues will be needed to clarify the situation.

\end{enumerate}

Our results show that galaxies that are more quenched tend to have longer molecular gas depletion times, suggesting that galaxy quenching is not only driven by exhaustion of gas supply but also a decline in star formation efficiency. In particular, our results are consistent with low SFE being the dominant driver for the stages from green valley to red galaxies, while the transition from main sequence to green valley is caused by a decrease of both molecular gas fraction and SFE. Therefore, it is likely that gas depletion is important in the initial stages of galaxy quenching, whereas a strong suppression of star formation efficiency then takes over to drive a deeper quenching into red galaxies.

\begin{acknowledgments}
\vspace{\baselineskip}

We thank S.-Y. Yu for helping with one of the GBT observing sessions. 
Y.-H.T. and A.D.B. acknowledge funding support from the National Science Foundation (NSF) under grant No. 2307441.
S.F.S. acknowledges the support by CBF-2025-I-236 project granted by the Secretaría de Ciencia, Humanidades, Tecnología e Innovación (SECIHTI) of
the Mexican Federal Government, and the PID2022-136598NB-C31 (ESTALLIDOS) grant by the Spanish Ministery of Science and Innovation (MCINN).
T.W. and K.D.F. acknowledge support from NSF grant 23-07440.
V.V. acknowledges support from the Comité ESO Mixto 2024 and from the ANID BASAL project FB210003. 
J.B-B acknowledges support from project UNAM DGAPA-PAPIIT AG 101025, Mexico and thanks the support from the PASPA 2025 grant.
Z.B. gratefully acknowledges the Collaborative Research Center 1601 (SFB 1601 sub-project B3) funded by the Deutsche Forschungsgemeinschaft (DFG, German Research Foundation) – 500700252.
R.H.-C. thanks the Max Planck Society for support under the Partner Group project "The Baryon Cycle in Galaxies" between the Max Planck for Extraterrestrial Physics and the Universidad de Concepción. R.H-C. also gratefully acknowledge financial support from ANID - MILENIO - NCN2024\_112 and ANID BASAL FB210003.
E.A.D.L. acknowledges the support of the PAPIIT-DGAPA IN100519 and IG100622 projects.
A.Z.L.A. gratefully acknowledges the support provided by the Postdoctoral Program (POSDOC) of UNAM (Universidad Nacional Autónoma de México). 
J.M.-L. acknowledges scholarship from ANID-Subdirección de Capital Humano/Doctorado Nacional/2023/21230541.

The National Radio Astronomy Observatory and Green Bank Observatory are facilities of the U.S. National Science Foundation operated under cooperative agreement by Associated Universities, Inc.
We acknowledge the usage of the SAO/NASA Astrophysics
Data System and the HyperLEDA database.

\end{acknowledgments}

\vspace{5mm}
\facilities{GBT}

\software{\texttt{spectral-cube} \citep{2019zndo...3558614G}, \texttt{matplotlib} \citep{Hunter:2007}, \texttt{numpy} \citep{harris2020array}, \texttt{scipy} \citep{Virtanen_2020}, \texttt{astropy} \citep{2022ApJ...935..167A}, \texttt{ipython} \citep{PER-GRA:2007}, \texttt{Pipe3D} \citep{2016RMxAA..52..171S,2016RMxAA..52...21S}, \texttt{reproject} \citep{2020ascl.soft11023R}, \texttt{edge-pydb} \citep{2024ApJS..271...35W}, \texttt{gbtpipe} (\url{https://github.com/GBTSpectroscopy/gbtpipe}), \texttt{degas} (\url{https://github.com/GBTSpectroscopy/degas}), \texttt{GBT-EDGE} (\url{https://github.com/teuben/GBT-EDGE}), \texttt{gbt-edge-analysis} (\url{https://github.com/ElthaTeng/gbt-edge-analysis}; \dataset[doi:10.5281/zenodo.20707910]{https://doi.org/10.5281/zenodo.20707910}) }

%\newpage

\appendix

\begin{figure}
\centering
\includegraphics[width=.8\linewidth]{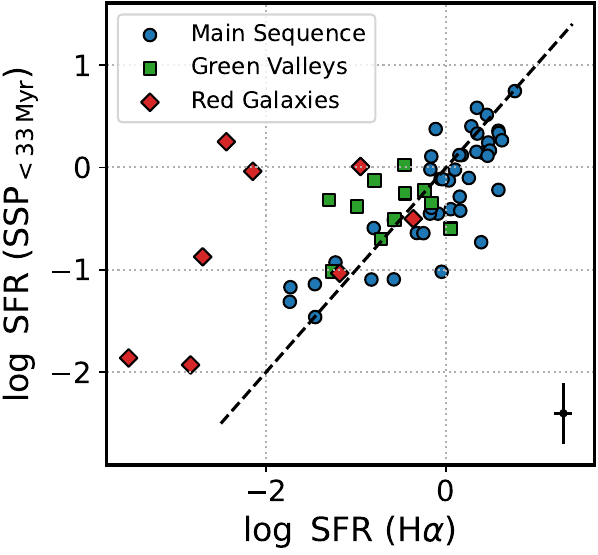}\\
\caption{Comparison between SFRs estimated via H$\alpha$ (Section~\ref{subsubsec:sfr}) and the simple stellar population (SSP) analysis. The dashed line indicates a one-to-one relation. For the MS and GV populations, SFRs inferred from H$\alpha$ are consistent with those inferred from the SSP analysis with $t < 33$~Myr. For RGs, however, H$\alpha$-based SFR generally provides better upper limits than using SSP. A typical error bar of $\sigma_x = \pm 0.1$ dex for SFR(H$\alpha$) and $\sigma_y = \pm 0.3$ dex for SFR(SSP) is shown in the lower right corner.}
\label{fig:sfr-compare}
\end{figure}

\begin{figure*}
\begin{minipage}{\linewidth}
\centering
\includegraphics[width=\linewidth]{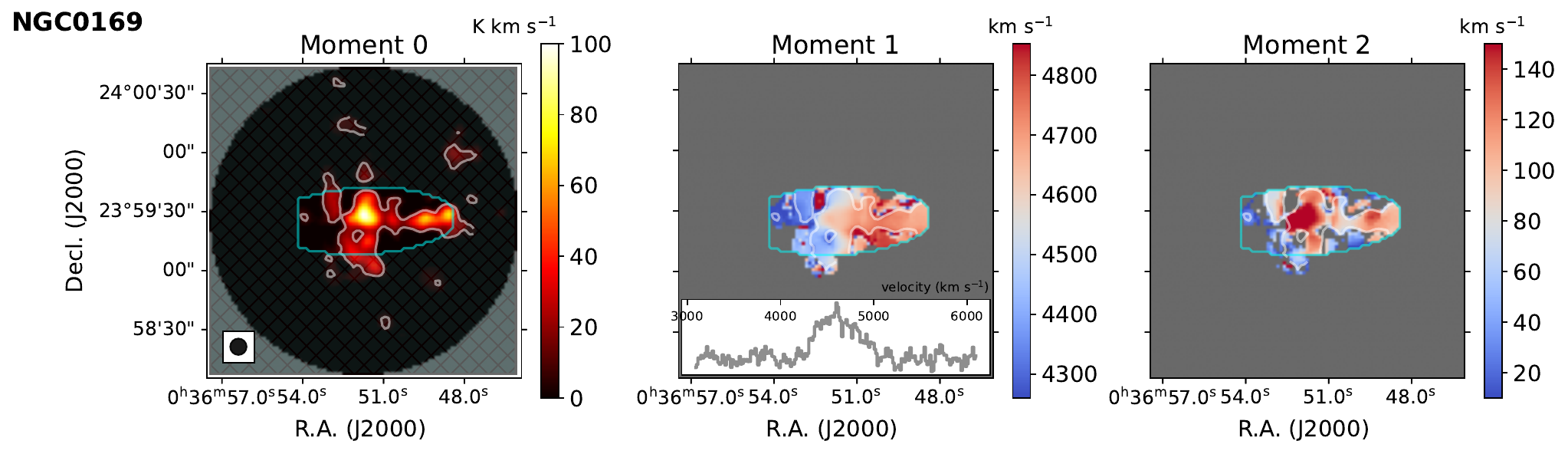}
\end{minipage}\\
\begin{minipage}{\linewidth}
\centering
\includegraphics[width=\linewidth]{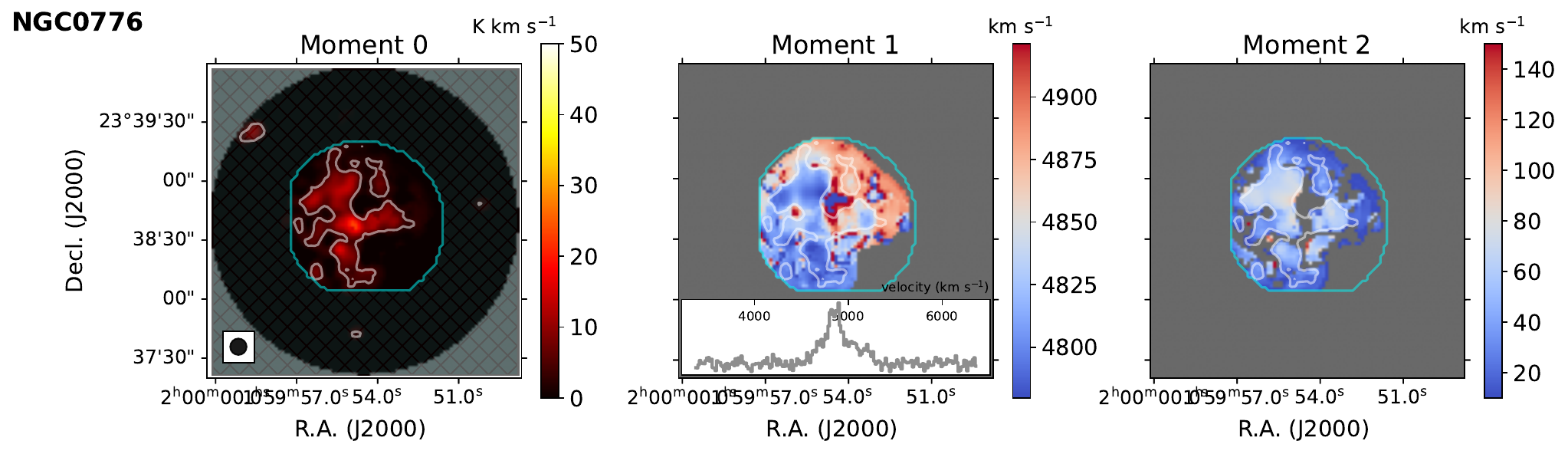}
\end{minipage}
\caption{Same as Figure~\ref{fig:mom0-datapref} but for all other galaxies in this work. The complete figure set (60 images) is available in the online journal.}
\label{fig:moments-datapref}
\end{figure*}

\begin{figure*}
\centering
\begin{minipage}{.33\linewidth}
\centering
\includegraphics[width=\linewidth]{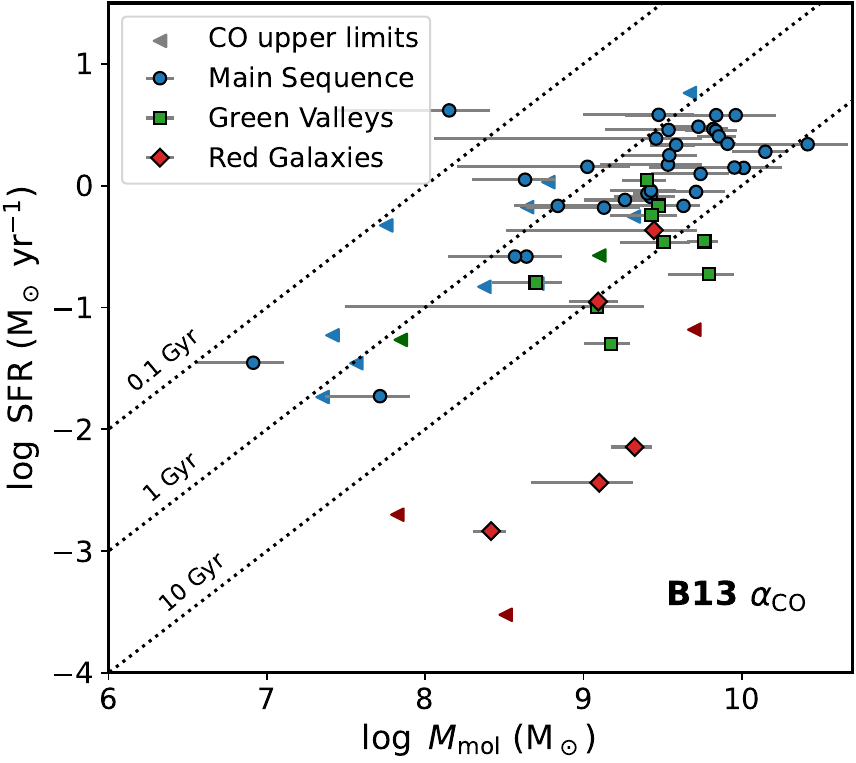}
\end{minipage}
\begin{minipage}{.315\linewidth}
\centering
\includegraphics[width=\linewidth]{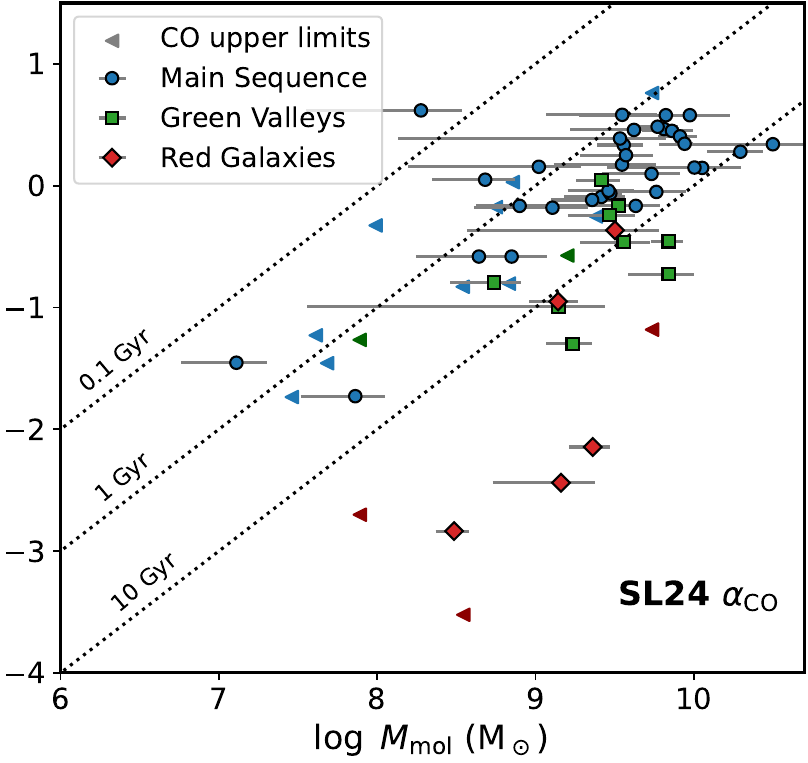}
\end{minipage}
\begin{minipage}{.315\linewidth}
\centering
\includegraphics[width=\linewidth]{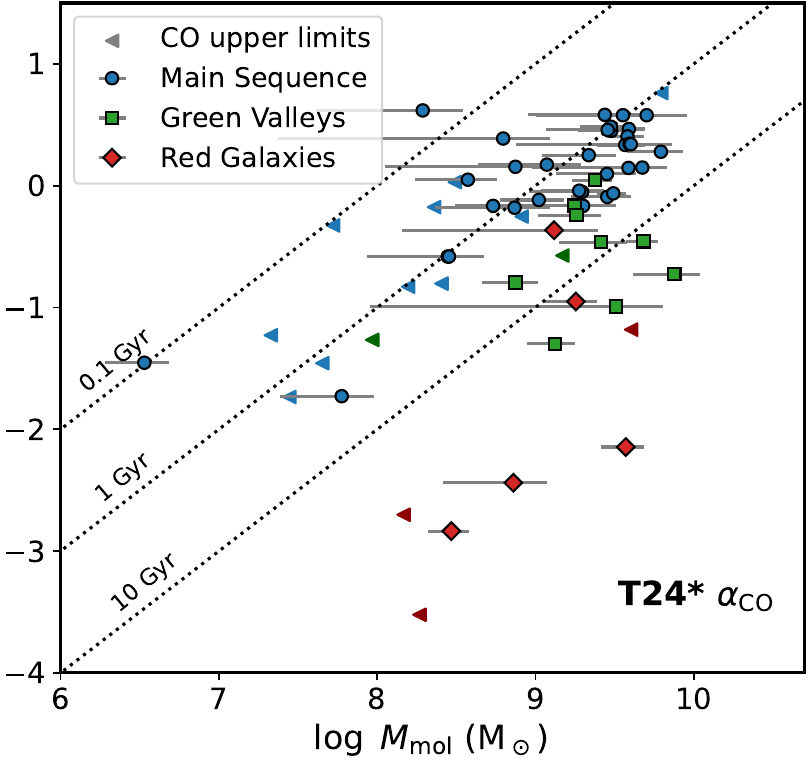}\\
\end{minipage}
\\
\begin{minipage}{.31\linewidth}
\centering
\includegraphics[width=\linewidth]{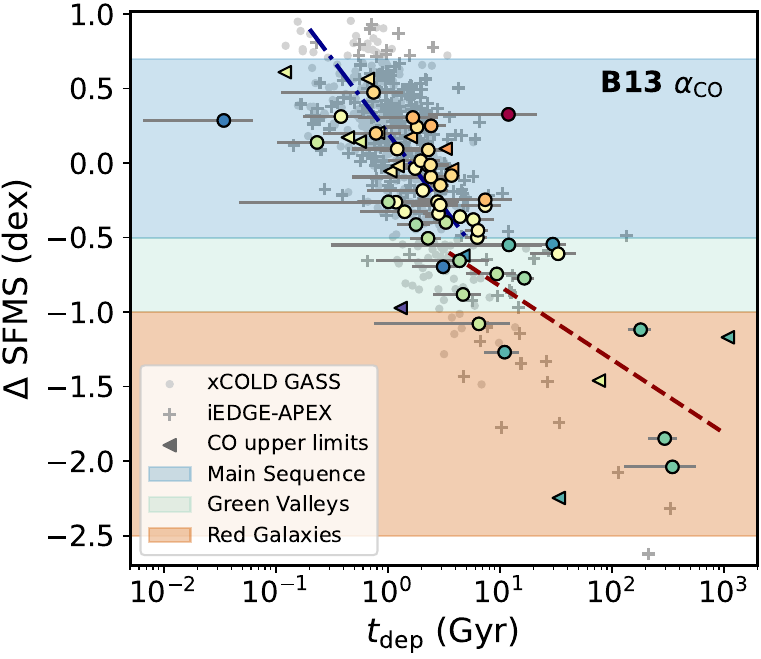}
\end{minipage}
%\hfill
\begin{minipage}{.295\linewidth}
\centering
\includegraphics[width=\linewidth]{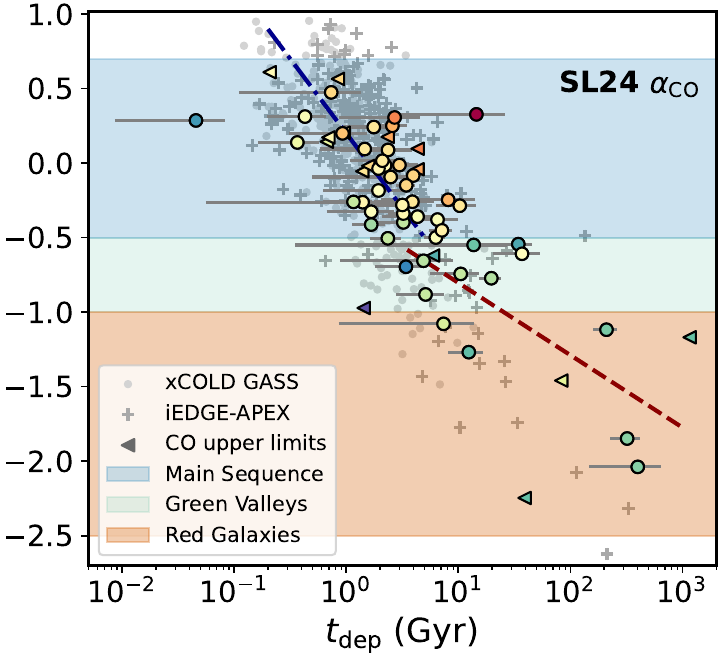}
\end{minipage}
%\hfill
\begin{minipage}{.37\linewidth}
\centering
\includegraphics[width=\linewidth]{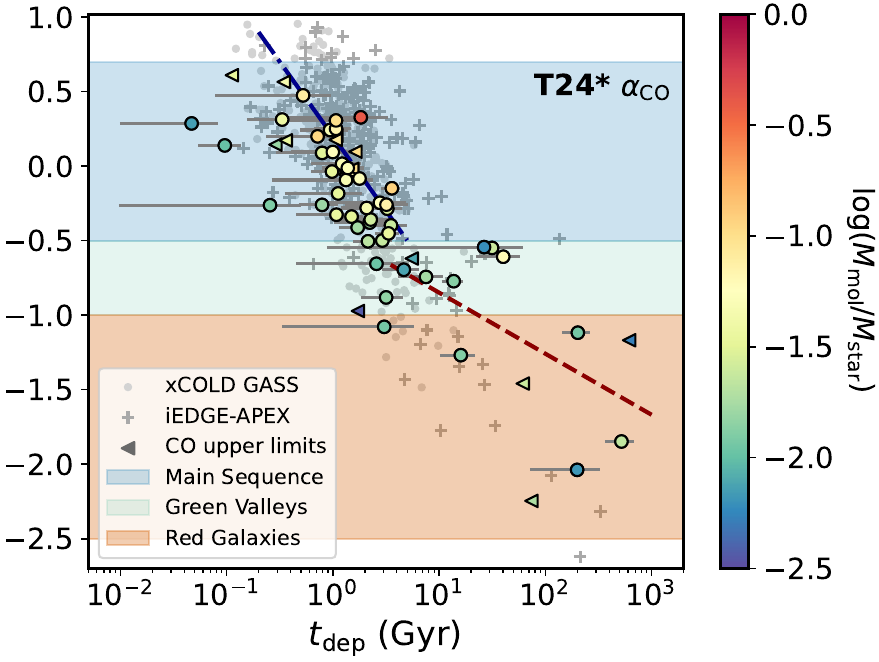}\\
\end{minipage}
\caption{Same as Figure~\ref{fig:scalings} but using $M_\mathrm{mol}$ and $t_\mathrm{dep}$ derived via the B13 (left), SL24 (middle), and T24* (right) \aco\ prescriptions. The best-fit slopes for quenched galaxies (red dashed lines) are -0.49, -0.48, and -0.41 for B13, SL24, and T24*, respectively. The qualitative results remain the same as using a constant \aco, as the \aco\ variation within and among prescriptions is small compared to the orders-of-magnitude span in $t_\mathrm{dep}$, and there is no systematic \aco\ dependence on $\Delta$SFMS.}
\label{fig:scalings_prescriptions}
\end{figure*}

\section{SFR Estimates from Simple Stellar Population Analysis} \label{appendix:ssp}

By leveraging the simple stellar population (SSP) and star formation history (SFH) tables in the CALIFA \texttt{Pipe3D} products, we compute the mass fraction of stars younger than 33~Myr ($f_{M\mathrm{young}}$)\footnote{Limited by the discrete age bins in the star formation history measurements from CALIFA, 33 Myr is used as the best approximation to a 30-Myr time scale.} within each galaxy and derive an integrated SFR via 
\begin{equation}
\mathrm{SFR}\ [\mathrm{M_\odot\ yr^{-1}}] = M_\mathrm{star,tot} \cdot f_{M\mathrm{young}}\ /\ 33\ \mathrm{Myr}\ ,
\label{eqn_sfr_ssp}
\end{equation}
where $M_\mathrm{star, tot}$ is the total stellar mass of a galaxy, and we define young stars as those younger than 33~Myr, following previous optical IFU surveys \citep{2016A&A...590A..44G,2021ApJ...909..131B,2022ApJS..262...36S}. 
These studies find that SSP-inferred SFRs using $t \lesssim 30$~Myr aligns best with the gold-standard SFRs obtained from the dust-corrected H$\alpha$ luminosity, while using a shorter timescale would lead to a systematic deviation. Indeed, we have checked on our sample that using e.g. $t < 12$~Myr for the SSP analysis generally increases the resulting SFRs by $\sim$0.2 dex, which is consistent with the finding in \citet{2022ApJS..262...36S}. As our goal here is to cross-compare SFR estimates from different tracers, we present only the SSP results using 33~Myr which align best with the H$\alpha$-based SFR estimates.

In detail, we derive the young mass fraction following these steps: 1) convert the luminosity fractions to mass fractions for each age-metallicity bin via SSP-computed mass-to-light ratios and the V-band image, 2) correct the pixel-by-pixel mass fractions for stellar dust attenuation via $10^{A_\mathrm{V}/2.5}$ using the $A_V$ map, 3) sum up all the masses contributed from age $<$ 33~Myr over all pixels within the $R_{25}$ radius, and 4) divide that mass contribution by the (also $A_\mathrm{V}$-corrected) integrated mass over all age-metallicity bins and within the $R_{25}$ radius.    

Figure~\ref{fig:sfr-compare} shows the comparison between the SSP- and H$\alpha$-derived SFRs for all galaxies in our sample. For MS and GV galaxies, we find consistent SFR estimates with both methods, which is in good agreement with previous studies \citep{2022ApJS..262...36S,2023MNRAS.526.5555S}. For RGs, however, it is clear that SSP-based SFRs are much higher than those using H$\alpha$, indicating even weaker upper limits than H$\alpha$-based SFRs. 
We consider the recombination line based SFRs more reliable and interpret this disagreement to indicate that the SSP analysis results in overestimates of SFRs in these galaxies.

Neither the SSP- nor the H$\alpha$-inferred SFRs are model independent, even though this fact is frequently overlooked. Both methods try to estimate the amount of stars, quantified by $M_\mathrm{star}$, that are formed in a recent time period, i.e., $\Delta M_\mathrm{star}$ in a certain $\Delta t$, as described before. In the case of the SSP-method this estimation is based on the derivation of the fraction of light corresponding to a population younger than a certain time-scale. Underlying this method it is assumed (i) a certain set of isochrones that traces the stellar evolution, (ii) an initial mass-function and the corresponding evolving mass-function,  (iii) an adopted stellar library to generate the SSP spectra and (iv) a dust-attenuation law and scenario (screen-model in our particular case). Modifying any of these assumptions would alter the derived SFR$_{\rm SSP}$. 

In many cases most of the weight in the discrepancy between both methods is placed on these assumptions, minimizing those required to derive the SFR based on H$\alpha$. Following the seminal exploration by \citet{1998ApJ...498..541K}, the SFR$_{\rm H\alpha}$ requires, in addition to the very same assumptions needed to evaluate SFR$_{\rm SSP}$, to assume (i) a particular shape for the recent star-formation history, (ii) the amount of ionizing photons produced by short-lived OB-stars ($t < 10$ Myr), that requires to assume a certain shape for their spectra in the highly uncertain far ultraviolet wavelength range, and (iii) a photionization model that links the number of these stars of a particular chemical composition with a certain H$\alpha$ luminosity. The number of physical parameters and assumptions regarding the ionized nebula behind the third assumption is indeed large and again broadly overlooked. 

As a result, any attempt to match both measurements in the literature has consistently provided with significantly strong, linear and almost slope-one correlations, both integrated galaxy-wide or spatially resolved down to $\sim$1~kpc for star-forming galaxies or regions \citep[][]{ARAA}. It has been also shown that in general there is a scaling factor between both quantities (an offset in logarithm scales). This indicates that (i) both methods are indeed somehow tracing the same physical quantity, i.e., the SFR, and (ii) all differences in the considered assumptions are translated somehow into a scaling factor that can be understood as time-scale renormalization: from $t\sim10$ Myr, in the case of H$\alpha$- to $t\sim33$ Myr, in the case of SSP-based SFR. However, it is important to remind that indeed we do not know the weight of each of these assumptions into this scaling factor or time-scale readjustment.

For this reason, the described offset between SFR$_{\rm SSP}$ and SFR$_{\rm H\alpha}$ and the highly overestimated SSP-based SFRs in RGs are likely due to a combination of reasons, and not a single one. First, previous results have reported the existence of a minimum threshold in the ability to recover a certain young stellar fraction when using similar stellar synthesis codes \citep[e.g., $\sim$3\%][]{2014A&A...562A..47G}. If this is the case, there would be a corresponding minimum SFR based on the SSP of 0.03$\times$M$_\mathrm{star,tot}$/33, that may affect the SFR derived for RGs (i.e., those with the lower values of $f_{M\mathrm{young}}$). Even if this limit is somehow overcome due to the most reliable recovery of low fractions of young stellar populations by \texttt{Pipe3D} \citep{2022NewA...9701895L}, there is still an expected mismatch with the values derived based on the H$\alpha$ luminosity: on one hand, a severe mask was applied to the spaxel-wise H$\alpha$ intensity (S/N$>$3), which for RGs may have removed a significant fraction of the spaxels in which the H$\alpha$ has a S/N of 1--2 \citep[e.g.,][]{2016A&A...588A..68G}; on the other hand, this mask was not applied to the SSP-based SFR estimates, and thus the integrated regions are also slightly different. 
Therefore, the reported SFR differences are expected based on the differences between the analyses, which affects more the RGs, i.e., those with low H$\alpha$ intensities and $f_{M\mathrm{young}}$.

\section{Additional Moment Maps} \label{appendix:maps}

While the analysis of this work is focused on global galaxy quantities, we present in Figure~\ref{fig:moments-datapref} the spatially resolved moment maps for all \Ngal galaxies in our GBT-EDGE sample. The color scale of the moment-0 maps represents a linear variation from 0 to 50~K~km~s$^{-1}$ for most galaxies, but to 100/200~K~km~s$^{-1}$ for NGC0169/CGCG536-030. For the moment-1 maps, the color scale ranges from the 5$^\mathrm{th}$ to 95$^\mathrm{th}$ percentile velocity values of each galaxy. The color scale of moment-2 maps ranges from 0 to 150 km~s$^{-1}$ for all galaxies. Regions outside the galaxy R$_{25}$ are excluded in the moment 1 and 2 maps to highlight velocity fields within the galaxies. 
A general description of these moment maps and the methods for map creation can be found in Section~\ref{subsec:maps}.

We caveat that regions outside the $R_{25}$ radii (cyan contours) may include spurious artifacts in the moment-0 maps. In our public data release, the masked moment maps that exclude regions outside $R_{25}$ are provided along with the reduced data cubes. The GBT-EDGE dataset is available on Zenodo under an open-source Creative Commons Attribution license: \dataset[doi:10.5281/zenodo.20707368]{https://doi.org/10.5281/zenodo.20707368}, and our analysis code is also available on GitHub and Zenodo: \dataset[doi:10.5281/zenodo.20707910]{https://doi.org/10.5281/zenodo.20707910}.

\section{Additional SFE Results with Varying \aco\ Prescriptions} \label{appdendix:sfe}

The SFR--$M_\mathrm{mol}$ relation and the relationship between $\Delta$SFMS and molecular gas depletion time ($t_\mathrm{dep}$) presented in Figure~\ref{fig:scalings} are based on a constant \aco\ assumption. Here we provide the same set of figures that use different \aco\ prescriptions, including B13 \citep{co-to-h2}, SL24 \citep{2024ARA&A..62..369S}, and T24* \citep[][and in preparation]{2024ApJ...961...42T} which are implemented and discussed in Section~\ref{sec:result}. The derived $M_\mathrm{mol}$, $t_\mathrm{dep}$, and their associated uncertainties based on these \aco\ prescriptions are all provided in our GitHub repository$^{\ref{gbt-edge-analysis}}$ (see Section~\ref{subsec:maps}). 

As shown by Figure~\ref{fig:scalings_prescriptions}, the systematic increase of $t_\mathrm{dep}$ from MS to GV and to RGs remain consistent regardless of the \aco\ choice. In particular, T24* even strengthens the anti-correlation across MS galaxies, showing the least scatter and exacerbating the $t_\mathrm{dep}$ deviation of GV and RGs from MS galaxies. The choice of \aco\ does not alter the $\Delta$SFMS--$t_\mathrm{dep}$ trend, since the \aco\ variations among these prescriptions are generally within a factor of two (see Table~\ref{tab:statistics} and Figure~\ref{fig:sfr-aco-compare}), which is almost negligible compared to the 2--3~dex range of $M_\mathrm{mol}$ and $t_\mathrm{dep}$ spanned by our galaxy sample. Therefore, we conclude that \aco\ variations are unlikely to have much impact on our results, while future \aco\ measurements across quenched galaxies will be needed for a solid verification.

\newpage

\bibliography{reference}{}

\bibliographystyle{aasjournalv7}



\end{document}